\documentclass[numbered appendix]{emulateapj}
\usepackage{longtable}

\slugcomment{}

\shorttitle{New Spatially Resolved Observations of the T Cha Transition Disk}
\shortauthors{Sallum et al.}

\begin{document}

\title{New Spatially Resolved Observations of the T Cha Transition Disk and Constraints on the Previously Claimed Substellar Companion}

\author{S. Sallum\altaffilmark{1}, J. A. Eisner\altaffilmark{1}, Laird M. Close\altaffilmark{1}, Philip M. Hinz\altaffilmark{1}, Andrew J. Skemer\altaffilmark{1}, Vanessa Bailey\altaffilmark{1}, Runa Briguglio\altaffilmark{2}, Katherine B. Follette\altaffilmark{1}, Jared R. Males\altaffilmark{1,4}, Katie M. Morzinski\altaffilmark{1,4}, Alfio Puglisi\altaffilmark{2}, Timothy J. Rodigas\altaffilmark{3}, Alycia J. Weinberger\altaffilmark{3}, Marco Xompero\altaffilmark{2}}
\altaffiltext{1}{Astronomy Department, University of Arizona, 933
  North Cherry Avenue, Tucson, AZ 85721, USA}
\altaffiltext{2}{INAF-Osservatorio Astrofisico di Arcetri, I-50125, Firenze, Italy}
\altaffiltext{3}{Carnegie Institution DTM, 5241 Broad Branch Rd, Washington, DC 20015, USA}
\altaffiltext{4}{NASA Sagan Fellow}

\email{email: ssallum@email.arizona.edu}

\begin{abstract}
We present multi-epoch non-redundant masking observations of the T Cha transition disk, taken at the VLT and Magellan in H, Ks, and L$'$ bands. T Cha is one of a small number of transition disks that host companion candidates discovered by high-resolution imaging techniques, with a putative companion at a position angle of 78$^\circ$, separation of 62 mas, and contrast of $\Delta L'$ = 5.1 mag. We find comparable binary parameters in our re-reduction of the initial detection images, and similar parameters in the 2011 L$'$, 2013 NaCo L$'$, and 2013 NaCo Ks data sets. We find a close-in companion signal in the 2012 NaCo L$'$ dataset that cannot be explained by orbital motion, and a non-detection in the 2013 MagAO/Clio2 L$'$ data. However, Monte-carlo simulations show that the best fits to the 2012 NaCo and 2013 MagAO/Clio2 followup data may be consistent with noise. There is also a significant probability of false non-detections in both of these data sets. We discuss physical scenarios that could cause the best fits, and argue that previous companion and scattering explanations are inconsistent with the results of the much larger dataset presented here.
\end{abstract}

\section{Introduction}\label{sec-intro}

Since their discovery, transition disks have been regarded as natural laboratories for the study of protoplanetary disk evolution and perhaps planet formation.  These objects' spectral energy distributions (SEDs) lack near- to mid-infrared emission, yet display a far-infrared excess \citep{strom89}.  Initial studies attributed these SED features to a lack of warm dust at inner, AU-scale radii, suggesting that they were ``in transition" from protoplanetary disks with excess throughout the infrared, to debris disks with only very weak far-infrared excess \citep[e.g.,][]{landp86,landp93,bryden99,calvet02}. More recent modeling of \emph{Spitzer} spectra \citep[e.g.,][]{calvet05, brown07,esp07a, esp07b, merin10} also associated the mid-infrared deficits with disk cavities or gaps on AU scales.  Followup submillimeter imaging \citep[e.g.,][]{brown09,hughes09,andrews11} has directly confirmed the presence of these features. 

Studies have identified several processes that could play a role in the formation of gaps and cavities, including photo-evaporative winds, grain growth, and dynamical interactions with companions. While photoevaporative winds would clear out only the gas and small dust in the inner disk, \citep{clarke01,alexander06}, accounting for radial drift of solids can lead to dissipation of dust at small radii as well \citep{alexandarm07}. Furthermore, X-ray winds may drive disk depletion at a faster rate than UV winds, suggesting that X-ray photoevaporation could clear our inner disk radii more efficiently \citep[e.g.,][]{ercolano08,drake09,owen10}. However, the cavity sizes and mass loss rates observed in transition disks are too large compared to their X-ray luminosities to be consistent with clearing by photoevaporation \citep{andrews11,owen11}. Furthermore, measurements of outer disk masses are too large compared to results of simulations that cause disks to go through an ``inner hole" phase \citep{alexander06}. While photoevaporation could explain some inner clearings, it alone cannot be responsible for transition disk structure.
 
Rather than clearing away disk material to lower the infrared emission, grain growth decreases its emissivity \citep{draine06,dalessio06}.  Growing grains to mm/cm sizes can create SED deficits comparable to those observed in transition disks \citep[e.g.,][]{dandd05, tanaka05,birnstiel12}. However, disk evolution simulations by Birnstiel et al. (2012) failed to generate the particle size distributions required to produce mm wavelength cavities. Additionally, soon after growing from $\sim1 \mu $m size to mm size, direct collisions between silicate particles could become destructive \citep[e.g.,][]{windmark12}, and the resulting smaller particles could then produce emission to fill in the SED deficit. This suggests that, while grain growth must impact disk evolution at some level, this process alone cannot shape transition disk cavities.

Dynamical interactions with companions are the best explanation to date for forming disk gaps. Models have demonstrated that stellar mass companions can open cavities in disks \citep[e.g.,][]{arty94,pichardo08}, and some observed disk gaps, such as those in CoKu Tau 4 \citep{ik08}, HD98800 \citep{furlan07}, and Hen 3-600 \citep{uchida04}, have been associated with stellar mass binaries. However, high resolution imaging has ruled out companions with masses higher than $\sim$ 20 - 30 M$_{Jup}$ for approximately half of the known transition disks \citep{k11, evans12}. This leaves the exciting possibility that planetary mass companions are clearing out cavities and gaps, accreting material that would have otherwise fallen onto the star \citep[e.g.,][]{najita07}. Simulations have shown that tidal interactions with a planetary mass companion can indeed open gaps in disks \citep[e.g.,][]{landp86,bryden99,crida06}. 

Here we discuss one transition disk thought to be shaped by a substellar mass companion, T Chamaeleontis (T Cha). T Cha is a G8 type, 1.5 M$_{\odot}$ star, first categorized as a weak-line T Tauri star due to its low H$\alpha$ equivalent width of $<10$ \AA~\citep{alc93}. This suggested it had entered the final stages of accretion. The classification conflicted with T Cha's infrared excess, thought to result from its outer disk; this spectral feature would place it as a classical T Tauri star.  Later observations showed that it is in fact a transition disk object, perhaps in the intermediate stages between a protoplanetary disk and a disk-free planetary system. \citet{brown07} found that T Cha's SED could be reproduced by a disk with a gap between 0.2 and 15 AU, and SED modeling by \citet{ol11} supported this, with a best fit gap extending from 0.17 to 7.5 AU. 

Imaging observations suggested the presence of a companion of L$'$ contrast 5.1 mag at a separation of 62 mas - 6.7 AU at a distance of 108 pc - \citep{hu11}, within T Cha's disk gap. However, \citet{olofsson13} showed that the detected signal could be modeled almost equally well by an asymmetry caused by forward scattering from the upper layers of the outer disk. Observing orbital motion of the companion candidate would distinguish between these two scenarios. To this end, we acquired new observations of T Cha with the Magellan AO (MagAO) system. We also present a re-analysis of the original discovery data from VLT/NaCo, as well as a new analysis of previously unpublished NaCo data from the Very Large Telescope (VLT) archive.

\section{Experimental Setup}\label{sec-exp}

\begin{deluxetable*}{lcccccc}
\centering
\tabletypesize{\scriptsize}
\tablecaption{Mask Parameters \label{tab-mpars}}
\tablewidth{0pt}
\tablehead{
\colhead{Mask} & \colhead{$N_{holes}$} & \colhead{$N_{b}$} & \colhead{$N_{cp}$} & \colhead{$N_{kp}$} & \colhead{Baseline Range (m)} & \colhead{Throughput (\%)}
}
\startdata
NaCo & 7 & 21 & 35 & 15 & 1.77 - 6.45 & 16\%\\
MagAO/Clio2 & 6 & 15 & 20 & 10 & 1.68 - 5.02 & 11\%
\enddata
\end{deluxetable*}

Non-redundant masking (NRM) transforms a filled aperture into a sparse interferometric array using a pupil-plane mask. While blocking the majority of the light, this provides much better knowledge of the PSF than a conventional telescope. A resulting image (called an interferogram) then shows the interference fringes formed by the mask, and subsequent image reconstruction or model fitting relies on quantities calculated from its Fourier transform. Since the mask is non-redundant (no two baselines have the same length and orientation), each baseline has unique $(u,v)$ coordinates. The symmetry of the Fourier transform means that identical information for a baseline can be found in two points in $(u,v)$ space - at $(u,v)$ and $(-u,-v)$. The finite size of the mask holes, as well as the width of the filter bandpass, causes this information to spread out. This means that the Fourier transform will have several distinct ``splodges," two coming from each mask baseline. 

Using the Fourier transform at the locations of these splodges, we find the complex visibility for each baseline, which has the form $Ae^{i \phi}$, where $A$ is the amplitude and $\phi$ the phase. Since atmospheric and instrumental effects corrupt the complex visibilities, we calculate two other quantities, squared visibilities and closure phases. Squared visibilities measure the power in the Fourier transform as a function of baseline length, and closure phases are sums of phases around three baselines that form a triangle. Closure phases eliminate atmospheric and instrumental phase offsets that corrupt measurements taken using single baselines. These obey the relation:
\begin{equation}
 \Phi_{cp} = \phi\left(u_1,v_1\right) + \phi\left(u_2,v_2\right) + \phi\left(u_3,v_3\right)
\end{equation}
where $u_i$ and $v_i$ are the sampling coordinates of the $i^{th}$ baseline in the Fourier plane. An N-hole mask will provide ${N \choose 2}$ baselines and visibilities, and ${N \choose 3}$ closure phases, ${N-1 \choose 2}$ of which are independent.

\section{Observations}\label{sec-obs}

\begin{deluxetable*}{lccccccccc}
\centering
\tabletypesize{\scriptsize}
\tablecaption{Summary of Observations \label{tbl-1}}
\tablewidth{0pt}
\tablehead{
\colhead{Target} & \colhead{Right Ascension} & \colhead{Declination}  & \colhead{t$_{int}$} & \colhead{N$_{frames}$}\tablenotemark{a} &
\colhead{N$_{visits}$}\tablenotemark{b} & \colhead{Total Time} & \colhead{Seeing} & \colhead{$\tau_0$}\\
\colhead{} & \colhead{(hh mm ss.sss)} & \colhead{(dd mm ss.sss)} & \colhead{(s)} & \colhead{} &\colhead{} & \colhead{(min)} & \colhead{(arcsec)} & \colhead{(ms)}
}
\startdata
\\
\multicolumn{9}{c}{\textbf{L$'$ Observations}}\\\\
\hline
\hline
\sidehead{\textbf{2010 Mar 17: VLT/NaCo}}
T Cha & 11 57 13.550 & -79 21 31.537 &  0.4 & 800 &  9 & 48 & 0.6 & 8\\
HD102260 & 11 45 13.822 & -78 36 58.633 & 0.4 & 800 & 10 & 53.3 & & \\
\hline
\sidehead{\textbf{2011 Mar 14: VLT/NaCo}}
T Cha & 11 57 13.550 & -79 21 31.537 & 0.4 & 800 & 7 & 37.3 & 2.0 & 1.5\\
HD102260 & 11 45 13.822 & -78 36 58.633 & 0.4 & 800  & 9 & 48 & & \\
\hline
\sidehead{\textbf{2012 Mar 8: VLT/NaCo}}
T Cha & 11 57 13.550 & -79 21 31.537 & 0.4 & 800 & 5 & 26.7 & 1.0 & 6\\
HD101251 & 11 37 49.220  & -79 14 31.604 & 0.4 & 800 & 2 &10.7 & & \\
HD102260 & 11 45 13.822 & -78 36 58.633 & 0.4 & 800 & 2 & 10.7 & & \\
\hline
\sidehead{\textbf{2013 Mar 25: VLT/NaCo}}
T Cha & 11 57 13.550 & -79 21 31.537 &  0.4 & 800 & 2 & 10.7 & 0.75 & 7 \\
T Cha & 11 57 13.550 & -79 21 31.537 &  0.3 & 1057 & 7 & 37.0 & & \\
T Cha & 11 57 13.550 & -79 21 31.537 &  0.2 & 1405 & 1 & 4.7 & & \\
HD102260 & 11 45 13.822 & -78 36 58.633 &  0.4 & 800 & 1 & 5.3 & &\\
HD102260 & 11 45 13.822 & -78 36 58.633 & 0.3 & 1057 & 4 & 21.1 & &\\
HD102260 & 11 45 13.822 & -78 36 58.633 & 0.5 & 1057 & 1 & 8.8 & &\\
HD101251 & 11 37 49.220  & -79 14 31.604 & 0.4 & 800 & 1 & 5.3 & &\\
HD101251 & 11 37 49.220  & -79 14 31.604 & 0.3 & 1058 & 2 & 10.6 & &\\
HD101251 & 11 37 49.220  & -79 14 31.604 & 0.15 & 1687 & 1 & 4.2 & & \\
\hline
\sidehead{\textbf{2013 Apr 5: Magellan/MagAO/Clio2}}
T Cha & 11 57 13.550 & -79 21 31.537 &  1.0 & 300 & 5 & 28.3 & 0.6 & N/A\\
T Cha & 11 57 13.550 & -79 21 31.537 &  1.0 & 200 & 1 & 28.3 & & \\
HD101251 & 11 37 49.220  & -79 14 31.604 & 0.9 & 300 & 2 & 9 & &\\
HD102260 & 11 45 13.822 & -78 36 58.633 & 1.3 & 300 & 1 & 15.16 & & \\
HD102260 & 11 45 13.822 & -78 36 58.633 & 1.3 & 200 & 2 & 15.16 & &\\
\hline
\hline\\
\multicolumn{9}{c}{\textbf{Ks Observations}}\\\\
\hline
\hline
\sidehead{\textbf{2010 Jul 1: VLT/NaCo}}
T Cha & 11 57 13.550 & -79 21 31.537 & 0.5 & 800 & 3 & 20 & 1 & 4\\
HD101251 & 11 37 49.220  & -79 14 31.604 & 0.5 & 800 & 1 & 6.7 & &\\
HD102260 & 11 45 13.822 & -78 36 58.633 & 0.5 & 800 & 1 & 6.7 & &\\
\hline
\sidehead{\textbf{2011 Mar 15: VLT/NaCo}}
T Cha & 11 57 13.550 & -79 21 31.537 & 2.0 & 200 & 6 & 40 & 0.75 & 3\\
HD102260 & 11 45 13.822 & -78 36 58.633 & 2.0 & 200 & 6 & 40 & &\\
\hline
\sidehead{\textbf{2013 Mar 26: VLT/NaCo}}
T Cha & 11 57 13.550 & -79 21 31.537 & 1.0 & 280 & 2 & 9.3 & 1 & 4 \\
T Cha & 11 57 13.550 & -79 21 31.537 & 0.8 & 427 & 4 & 22.8 & &\\
T Cha & 11 57 13.550 & -79 21 31.537  & 0.7 & 497 & 3 &17.4 & &\\
T Cha & 11 57 13.550 & -79 21 31.537  & 0.5 & 807 & 2 & 13.5 & &\\
HD102260 & 11 45 13.822 & -78 36 58.633 & 1.0 & 350 & 1 & 5.8 & &\\
HD102260 & 11 45 13.822 & -78 36 58.633 & 0.5 & 707 & 2 & 11.8 & &\\
HD102260 & 11 45 13.822 & -78 36 58.633  & 0.4 & 707 & 1 & 4.7 & &\\
HD102260 & 11 45 13.822 & -78 36 58.633 & 0.3 & 1057 & 1 & 5.3 & &\\
HD101251 & 11 37 49.220  & -79 14 31.604  & 0.2 & 1407 & 3 & 14.1 & &\\
HD101251 & 11 37 49.220  & -79 14 31.604  & 0.11 & 2782 & 1 & 5.1 & &\\
\hline
\hline\\
\multicolumn{9}{c}{\textbf{H Observations}}\\\\
\hline
\hline
\sidehead{\textbf{2013 Mar 27: VLT/NaCo}}
T Cha & 11 57 13.550 & -79 21 31.537 & 0.4 & 605 & 1 & 4.03 & 0.75 & 3.5\\
T Cha & 11 57 13.550 & -79 21 31.537 & 0.4 & 847 & 1 & 5.64 & &\\
T Cha & 11 57 13.550 & -79 21 31.537 & 0.5 & 707 & 4 & 23.6 & &\\
HD102260 & 11 45 13.822 & -78 36 58.633 & 0.15 & 2107 & 3 & 15.8 & &\\
HD101251 & 11 37 49.220  & -79 14 31.604 & 0.11 & 3157 & 2 & 11.6 & &\\
HD101251 & 11 37 49.220  & -79 14 31.604 & 0.11 & 1353 & 1 & 2.5 & &
\enddata
\tablenotetext{a}{Number of frames in each visit}
\tablenotetext{b}{Each visit consists of all images taken before switching between target and calibrator.}
\label{tab-obs}
\end{deluxetable*}

\begin{figure*}
\epsscale{1.2}
\plotone{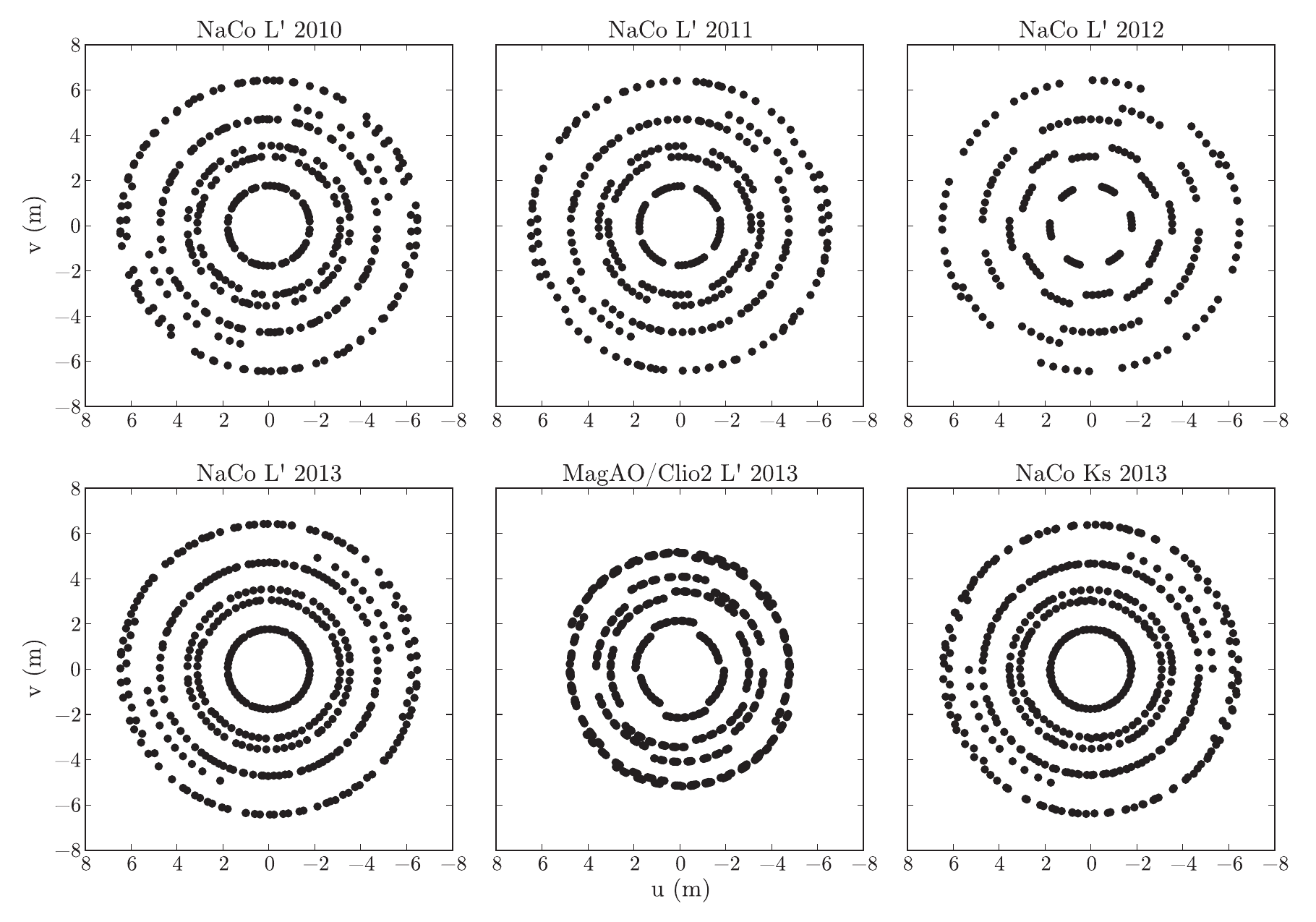}
\caption{Sky rotation comparison in $(u,v)$ space for all L$'$ data sets and 2013 NaCo Ks data. The NaCo 2010 L$'$ data had a change in sky rotation of $\sim62^\circ$. The total sky rotation in the NaCo 2011 L$'$ data was $\sim39^\circ$. The NaCo 2012 L$'$ data had $\sim19^\circ$ of sky rotation, and the NaCo 2013 L$'$ data had $\sim56^\circ$. The MagAO/Clio2 2013 data had a change of $\sim47^\circ$. Due to its smaller aperture, the MagAO/Clio2 baselines are shorter than for NaCo. Lastly, the NaCo 2013 Ks data had $\sim71^\circ$ of sky rotation.}
\label{fig-good_rot}
\end{figure*}

\subsection{New Magellan/MagAO/Clio2 Data}

We observed T Cha and two unresolved calibrators, HD101251 and HD102260 using the 6.5 m Clay telescope, MagAO adaptive optics system \citep{close13,morzinski14}, and Clio2 science camera \citep{freed04,sivana06} on 2013 April 5. A six-hole non-redundant mask  was mounted in a pupil plane filter wheel in Clio2. Table \ref{tab-mpars} lists the parameters of the Clio2 mask. We used two calibrators to lessen the probability that detected signals were being injected by a resolved or binary calibrator. Our exposure times for T Cha, HD101251, and HD102260 were 1.0 s, 0.9 s, and 1.3 s, respectively. We took 4 to 6 50-frame data cubes (called ``visits" in Table \ref{tab-obs}) before switching objects according to the pattern target-cal1-target-cal2. We acquired 6 visits for T Cha, 2 visits for HD101251, and 3 visits for HD102260. These resulted in 28.3 minutes, 9 minutes, and 15.16 minutes of total integration, respectively. We observed in L$'$, $\lambda_{c} = 3.78$ $\mu$m, and the total change in sky rotation angle was $47^\circ$, shown in Figure \ref{fig-good_rot}.

\subsection{Previously Published 2010 VLT/NaCo Data}

\subsubsection{2010 March 17 L$'$ Observations}

The initial detection of the T Cha companion \citep{hu11} resulted from NRM observations taken at the VLT using NaCo \citep{penzen93,rousset03}. The parameters for the NaCo mask are listed in Table \ref{tab-mpars}. In order to verify our reduction pipeline, we downloaded and re-reduced the previously published observations from the archive. This dataset included observations of T Cha and a single, unresolved calibrator, HD102260, with 9 visits to T Cha and 10 to HD102260. The observations followed the pattern ...cal-target-cal..., dithering so that each 100-frame data cube placed the interferogram on one of the detector's quadrants. These data resulted in 48 minutes of integration on T Cha and 53.33 minutes on HD102260. Images were taken in L$'$, $\lambda_{c} = 3.8$ $\mu$m, and the total change in sky rotation angle was $62^\circ$, shown in Figure \ref{fig-good_rot}.

\subsubsection{2010 July 1 Ks Observations}

The detection of T Cha in L$'$ was accompanied by a non-detection in Ks ($\lambda_c = 2.18\mu$m), from data taken at the VLT using NaCo in July 2010. We reduced this archival dataset, which includes observations of T Cha, HD102260, and HD101251. Three visits were made to T Cha, totaling 20 minutes of integration. For each calibrator, 1 visit consisting of 6.7 minutes of integration was made. The change in sky rotation angle for these data was 19$^\circ$, and the dithering pattern was identical to the 2010 L$'$ NaCo data. These data were too noisy to detect companion signals comparable to the Huelamo et al. (2011) binary, so we include their discussion in Appendix \ref{app-otherdata}.

\subsection{Unpublished VLT/NaCo Data}
While searching for published data to verify our pipeline, we found additional observations taken using VLT/NaCo from 2011 - 2013. These include L$'$ and Ks observations from March 14-15, 2011, L$'$ observations from March 8, 2012, and L$'$, Ks, and H band observations from March 25-27, 2013. We include a description of each dataset in this section. However, the scatter in the 2011 Ks and 2013 H band observations would wash out companion signals of interest. For this reason, we include only a short discussion of the results from these data in Appendix \ref{app-otherdata}. 

\subsubsection{2011 March 14 L$'$ Observations}

We reduced archival L$'$ VLT/NaCo data taken in 2011, which included observations of T Cha and the same calibrator as the 2010 L$'$ dataset, HD102260. These data consisted of seven visits to T Cha, and nine visits to HD102260, resulting in 37.33 minutes of integration on T Cha and 48 minutes on HD102260. The dithering pattern was the same as for the 2010 L$'$ NaCo data, and the total change in sky rotation angle was $39^\circ$, shown in Figure \ref{fig-good_rot}.

\subsubsection{2011 March 15 Ks Observations}

These archival data include observations of T Cha and HD102260 taken in Ks. The dithering pattern was identical to the 2010 L$'$ NaCo data. Six visits were made to each object, resulting in 40 minutes on target and calibrator. The total change in sky rotation for these data was $38^\circ$, shown in Appendix \ref{app-otherdata}.

\subsubsection{2012 March 8 L$'$ Observations}

We reduced archival VLT/NaCo L$'$ data taken on March 8, 2012. These include observations of T Cha, HD102260, and HD101251. Five visits were made to T Cha and 2 to each calibrator. This resulted in a total of 26.7 minutes of integration for T Cha, and 10.7 minutes for each calibrator. The visits were dithered so that the image fell on a different detector quadrant during neighboring sets of 100 exposures. The total change in sky rotation for this dataset was 19$^\circ$, shown in Figure \ref{fig-good_rot}.

\subsubsection{2013 March 25 L$'$ Observations}

The 2013 archival L$'$ VLT/NaCo data include observations of T Cha, HD101251, and HD102260. Ten visits were made to T Cha, with 52.3 total minutes of integration.  A total of 4 visits were made to HD101251, resulting in 20.1 minutes of integration, and the 6 visits to HD102260 yielded 35.3 minutes of integration time. These data had a change in sky rotation of 56$^\circ$, shown in Figure \ref{fig-good_rot}. Table \ref{tab-obs} details the individual visits to all three objects, which did not have identical exposure times.

\subsubsection{2013 March 26 Ks Observations}

We also present archival VLT/NaCo data taken in Ks band. The 11 visits to T Cha resulted in a total of 63 minutes of integration. HD101251's 4 visits yielded 19.2 minutes of total exposure time, and the total amount of integration for the 5 visits made to HD102260 was 26 minutes. The change in sky rotation for this dataset was 71$^\circ$, shown in Figure \ref{fig-good_rot}. See Table \ref{tab-obs} for details of the individual visits to each object.

\subsubsection{2013 March 27 H Observations}

The last archival dataset was taken on March 27, 2013 in H band ($\lambda_c = 1.65 \mu$m). The total integration time for T Cha's 6 visits was 33.3 minutes. For HD101251, 2 visits were made, resulting in 14.1 minutes of integration. Lastly, the 3 visits to HD102260 total 15.8 minutes of integration time. Table \ref{tab-obs} details the individual visits made to each object. The change in sky rotation for these observations was 39$^\circ$, shown in Appendix \ref{app-otherdata}.

\section{Data Reduction}\label{sec-red}

\begin{figure}
\epsscale{1.2}
\plotone{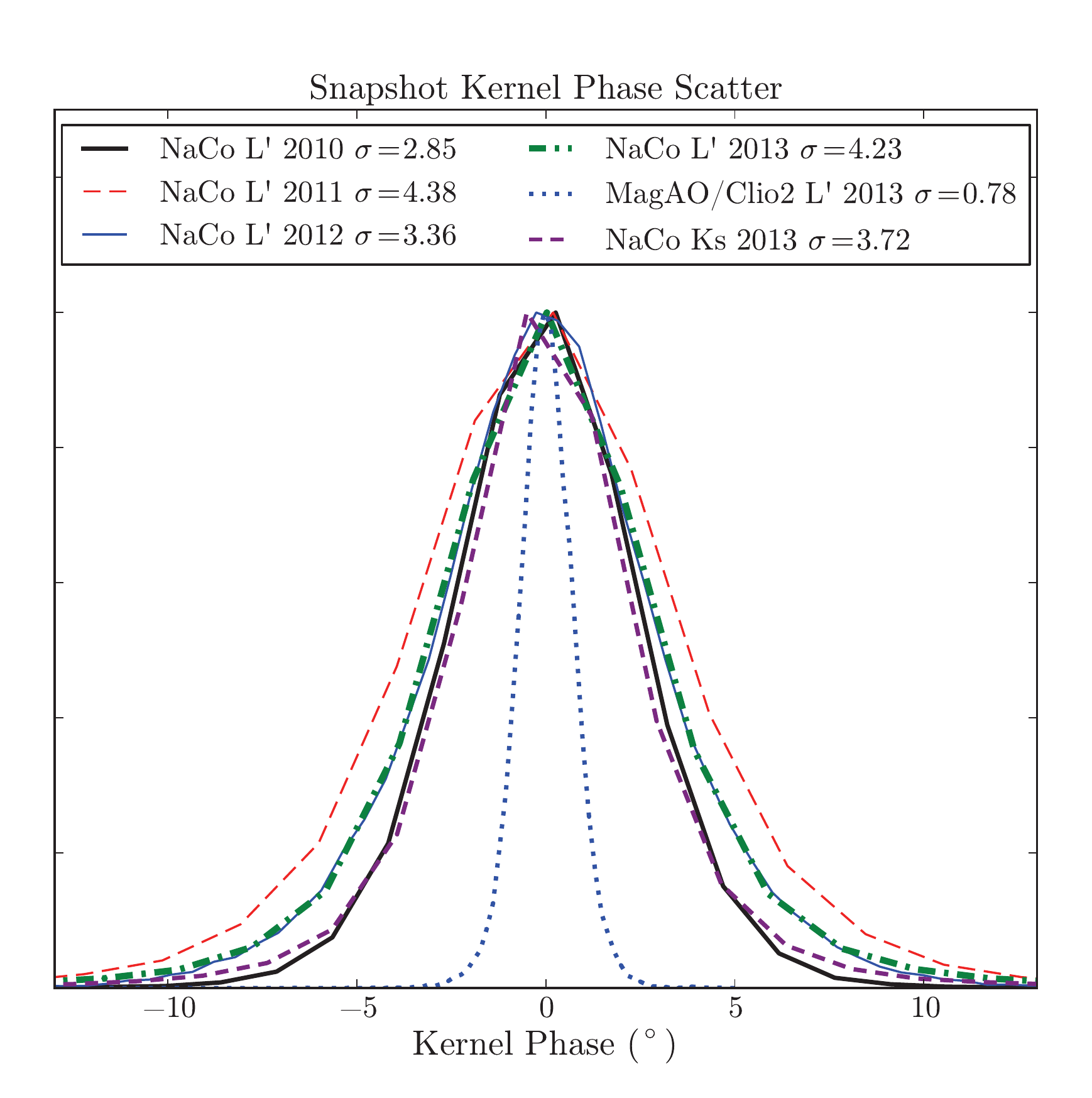}
\caption{Normalized histograms of uncalibrated kernel phases for L$'$ and 2013 Ks data. For a subset of each dither (chosen so that equal amounts of integration came from all observations), we subtract the mean kernel phase from each individual measurement to generate the histograms shown above. The snapshot kernel phase errors are much lower in the MagAO/Clio2 data than in the NaCo data sets, indicating lower levels of random noise.}.\label{fig-good_snapshot}
\end{figure}

\begin{figure*}
\epsscale{1.1}
\plotone{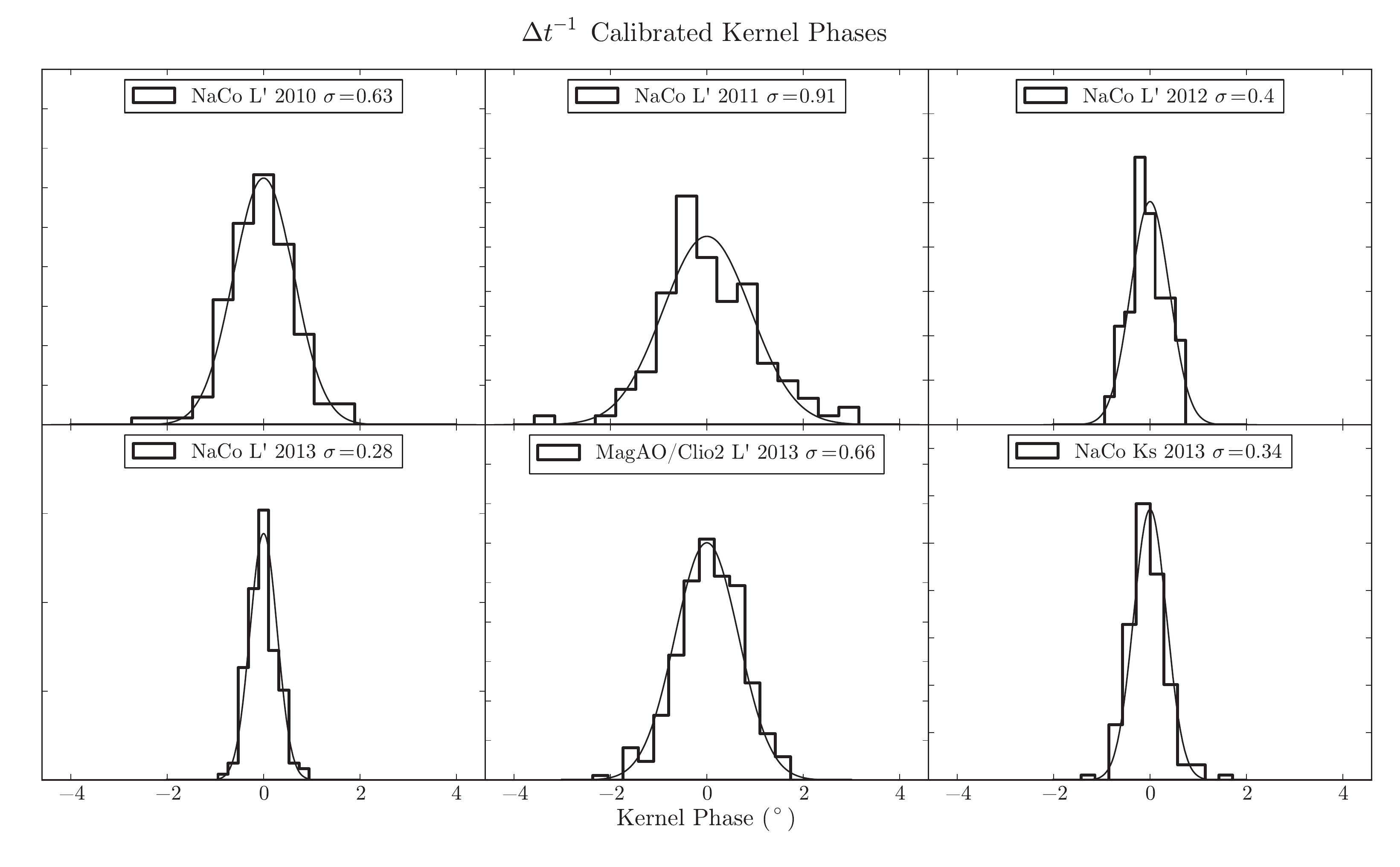}
\caption{Normalized histograms of each set of calibrated L$'$ kernel phases as well as NaCo 2013 Ks kernel phases, with their best fit Gaussian distributions over plotted. The Gaussian distributions were used to generate noise realizations for the simulations described in Section \ref{sec-sims}.}.\label{fig-goodcal}
\end{figure*}

We have developed a suite of software in Python to perform basic data reduction, calibration, and visibility and closure phase calculations. We first flat-field and bad-pixel correct all images. For a given set of two dithers, we perform sky subtraction for one position by taking the median of all images in the other dither position and subtracting the median from each image. We then apply a super-Gaussian window function to reduce the noise associated with the low-signal edges of the interferogram and spatially filter the data. A super-Gaussian has the form exp$(-kx^4)$; we choose $k$ such that the half width at half max is $\lambda / d_{sub}$, where $d_{sub}$ is the mask sub-aperture diameter \citep[e.g.,][]{bernatphd}. After windowing, we Fourier transform the data.

Next, we calculate squared visibilities. We first simulate an interferogram and resulting Fourier transform using the locations and sizes of the holes in the mask, along with the observation's wavelength. This provides us with the pixel locations of the splodges (see $\S$ \ref{sec-exp}) in the Fourier transform. We then square the Fourier transform of the data, and sum all pixels within the splodges corresponding to each baseline, normalizing by the total power in the interferogram. 

The typical uncertainty due to random errors for the visibilities ranges between 0.02 and 0.12 for all L$'$ and 2013 Ks data sets. For comparison, a binary with separation 62 mas and contrast $\Delta L' = 5.1$ mag produces a change in visibility between the shortest and longest baseline of approximately 0.035 for both the MagAO/Clio2 and NaCo masks. While some of the followup data sets' random visibility errors are smaller than this signal, visibilities' dependence on the AO system \citep[e.g.,][]{lacour11,ki12} renders them harder to calibrate. Differences in AO performance over the night, or between target and calibrator observations can introduce additional error. To investigate this, we divide the visibilities for each set of two adjacent calibrator scans; the calibrated visibilities for a point source should be equal to 1 for all baselines. We then take the scatter in these calibrated visibilities as an estimate for the systematic uncertainties in the target visibilities. This results in calibrated visibilities with scatters ranging between 0.04 and 0.11. Due to these large uncertainties, as in previous NRM studies \citep[e.g.,][]{hu11,ki12} we restrict our binary fitting to phases only, rather than including the squared visibilities. 

For each triangle, we form the bispectrum by multiplying the complex values in the Fourier transform at three $\left(u,v\right)$ coordinates. We then average over the individual frames, and take the phase of the average bispectrum to be our closure phase. Of the ${N \choose 3}$ closure phases, only ${N-1 \choose 2}$ are independent.  For this reason, we perform fits on kernel phases, linearly independent combinations of closure phases (see Martinache 2010, Kraus \& Ireland 2012, and Ireland 2013).  We find our kernel phases in a way similar to \citet{m10}. Appendix \ref{app-kpproj} gives a detailed description of our projection method. 

Figure \ref{fig-good_snapshot} shows histograms of the kernel phases for individual images in the Ks and L$'$ data sets, and can be taken to represent a comparison of the snapshot kernel phase errors for the different data sets. For a given dither, we calculate the ${N-1 \choose 2}$ mean kernel phases. We then calculate the kernel phases for every individual image in the dither and subtract the mean. We use the same total integration time, 39 s, to calculate the mean kernel phase. This process yields the distributions shown in Figure \ref{fig-good_snapshot}. The snapshot errors in the 2013 MagAO data ($\sigma = 0.73 ^\circ$) are much smaller than the VLT/NaCo data sets ($\sigma = 2.85 - 4.38^\circ$ for L$'$ kernel phases.) We also compare the MagAO/Clio2 and NaCo data with different amounts of time-averaging.  As long as these averages are performed within a single dithering sequence, the chosen interval does not change the noise levels significantly. Due to their high scatter and the resulting unreliable fits we show kernel phase histograms for the 2010 and 2011 Ks as well as the 2013 H band data in Appendix \ref{app-otherdata}. 

Random sources of noise associated with both AO performance and observing conditions cause snapshot kernel phase errors. For exposures much longer than the inverse of the AO system bandwidth, closure phase errors scale in the following way \citep{i13}:
\begin{equation}
\sigma_{cp} \sim \sigma_{\phi} \sqrt{\frac{1}{f_c T}}~,
\end{equation}
where $\sigma_\phi$ is the phase noise on each sub-aperture in the closing triangle, $f_c$ the cutoff frequency below which piston noise is white, and T the exposure time. The kernel phase errors will then be the projection of the closure phase errors. If we assume $\sigma_\phi$ and $f_c$ are equal for all data sets, the ratio of two observations' closure phase errors scales with the ratio of their exposure times. Using 0.2s -- 0.5s for the NaCo data, and 1.0s for the MagAO/Clio2 data, the ratio of the MagAO/Clio2 to NaCo closure phase errors should range from 0.44 to 0.71. The ratio of the MagAO/Clio2 and NaCo L$'$ observed errors ranges between 0.27 and 0.18 depending on the NaCo L$'$ dataset. Hence, exposure time alone cannot explain the discrepancy between the MagAO/Clio2 and NaCo L$'$ snapshot errors. 

Other random sources of noise include photon, background, and read noise, which lead to closure phase errors \citep[following][]{i13}:
\begin{equation}
\sigma_{cp} \sim \frac{N_h}{N_p V}\sqrt{1.5(N_p+N_b+n_p \sigma^2_{ro})}~,
\end{equation}
where $N_h$ is the number of holes in the mask, $V$ the fringe visibility, $N_p$ the total number of photons, $N_b$ the number of background photons, $n_p$ the number of pixels, and $\sigma_{ro}$ the read noise. If we assume we are in the photon-noise regime, the closure phase error simplifies to
\begin{equation}
\sigma_{cp} \sim \frac{N_h}{V}\sqrt{\frac{1.5}{Np}}~.
\end{equation}
The number of photons is $N_p = F_{target} \times A_{tel} \times f_{mask} \times T$, where $F_{target}$ is the photon flux from the source, $A_{tel}$ the telescope collecting area, $f_{mask}$ the fraction of light allowed through by the mask, and $T$ the exposure time. The relevant values for the MagAO/Clio2 observations are $A_{tel} \sim (6.5$m$)^2$, $f_{mask} = 0.11$, $T = 1s$, and $N_h = 6$. For the NaCo observations, $A_{tel} \sim (8.2$m$)^2$, $f_{mask} = 0.16$, $T = 0.2s-0.4s$, and $N_h = 7$. Assuming that the fringe visibilities for the two data sets are approximately equal, taking the ratio of the MagAO/Clio2 to NaCo closure phase errors gives $0.58-0.92$. Thus, the differences in telescope and observing parameters cannot fully explain the lower MagAO/Clio2 snapshot errors. Better AO performance by MagAO, which leads to lower values for $\sigma_{\phi}$ and higher values of $f_c$, could be one cause for this discrepancy. Additionally, the smaller holes (as evinced by the lower throughput in Table \ref{tab-mpars}) of the Clio2 mask mean that the 2013 observations are less redundant than the NaCo observations. This could also reduce the snapshot errors for the MagAO/Clio2 data.

\section{Calibration}\label{sec-calib}
Due to atmospheric and instrumental systematics, the mean kernel phases themselves (subtracted off in Figure \ref{fig-good_snapshot}) can vary substantially throughout the observations. To take these effects into account, we subtract our unresolved calibrator kernel phases from our target kernel phases.  We do this in several ways. 

To apply a simple nearest-neighbor calibration, we first calculate the time between a given target scan and all calibrator scans ($\Delta t$). We then average all calibrator scans, weighting by $\Delta t^{-10}$, and subtract the weighted-average calibrator from the target scan. To use information from all of the calibrator scans, rather than limiting ourselves to only the nearest-neighbor, we calculate an average calibrator, weighting the scans by $\Delta t^{-1}$. Finally, we apply a more optimized weighting, similar to LOCI \citep{loci} techniques in direct imaging data reduction and the calibration strategy adopted in Kraus \& Ireland (2012) and Ireland (2013). 

For the LOCI-like calibration scheme, we find the linear combination of calibrator scans that minimizes the sum of the target's squared kernel phases. This is equivalent to minimizing the $\chi^2$ for the null model. While this calibration scheme provides the lowest scatter and thus highest signal to noise, it can also subtract signal from the measurements. For this reason, it is often applied iteratively, minimizing the $\chi^2$ for the null model initially and then minimizing the $\chi^2$ for the best-fit model until the best-fit converges. We also LOCI calibrate without iteration, minimizing the $\chi^2$ of the $\Delta t^{-1}$ model. Both LOCI schemes remove signal, and do not always give consistent results. Therefore, we focus on the fits to simpler, neighbor-like calibrations in the subsequent sections. The behavior of the LOCI calibration method will be detailed in a future paper.

The histograms for the L$'$ and 2013 Ks data sets are shown in Figure \ref{fig-goodcal}. While the MagAO/Clio2 data have lower snapshot kernel phase errors than the NaCo L$'$ observations (see Figure \ref{fig-good_snapshot}), the systematics in the three data sets are such that the scatter in the mean, calibrated kernel phases are quite similar. We speculate that the greater temporal spacing between target and calibrator observations in the MagAO/Clio2 observations could cause this. The average time between target and calibrator scans in the these data is 8.8 minutes. For the NaCo observations, the mean time between target and calibrator scans for the L$'$ data ranges between 4.2 and 6.9 minutes. 

\subsection{Consistency Checks}

To check whether one calibrator could be contaminating the MagAO/Clio2 data and increasing the kernel phase scatter, we calibrated our T Cha kernel phases using each calibrator individually. Our dithering pattern alternated calibrator observations between target observations, resulting in some target scans being closer in time to one of the two calibrators. Using only one calibrator star increased the scatter in the calibrated kernel phases. However, both single-star calibrations yielded kernel phases with nearly identical standard deviations (approximately 1$^\circ$). This similarity suggests that neither calibrator is contaminating the dataset, and the increase in noise compared to the two-star calibration scheme highlights the need for calibrator scans taken close in time to the target observations. Furthermore, best fits to these calibrated data are consistent with those for the fully calibrated data.

For the fully-calibrated MagAO/Clio2 L$'$ data, we compare the scatter in scans taken during the first half of the night to the scatter for those taken during the second. The kernel phase standard deviations in this test are nearly identical (0.66$^\circ$ and 0.68$^\circ$). This suggests that the calibration quality did not change significantly during the observations.

We carried out calibrator tests for the NaCo 2012 L$'$ and 2013 L$'$ and Ks data sets, which, unlike the 2010 and 2011 data, included observations of two calibrator stars. We calibrated each calibrator star using the other with a $\Delta t^{-1}$ method, and then fit a binary model to the kernel phases. In all three data sets, neither calibrator star's kernel phases show clear signs of a companion; simulations show that their best fits are consistent with noise.

We also calibrated the NaCo 2012 and 2013 T Cha kernel phases using each calibrator separately, and then fit the resulting kernel phases. Using individual calibrators did not change the best fits significantly for any of these three data sets. Additionally, the scatters for these calibrations were similar. Using only HD101251 and HD102260, respectively, the standard deviations were 0.39$^\circ$ and 0.36$^\circ$ for 2012 L$'$, 0.28$^\circ$ and 0.33$^\circ$ for 2013 L$'$, and 0.45$^\circ$ and 0.36$^\circ$ for 2013 Ks.

We also check that the scatters in these three NaCo data sets are comparable for both halves of the night. The 2012 L$'$ data from the first and second halves of the observations have scatters of 0.37$^\circ$ and 0.36$^\circ$, respectively. For the 2013 L$'$ observations, the kernel phase standard deviation for the first half of observations is 0.32$^\circ$, while for the second half it is 0.27$^\circ$. The change throughout the night for the 2013 Ks observations was slightly larger, with a standard deviation of 0.45$^\circ$ for the first half, and 0.31$^\circ$ for the second half. 

Lastly, we compare the scatter in the calibrated kernel phases (Figure \ref{fig-goodcal}) with changes in average seeing and coherence time (Table \ref{tab-obs}) between observations. In general, observations with longer coherence times and lower wind speeds had lower kernel phase scatter.

\section{Binary Fitting}\label{sec-fitting}

\begin{deluxetable}{lcccc}
\tabletypesize{\scriptsize}
\tablecaption{Binary Grid Parameter Space \label{tab-bgrid}}
\tablewidth{0pt}
\tablehead{
\colhead{Parameter} & \colhead{Minimum} & \colhead{Maximum} & \colhead{Step Size} \\
}
\startdata
P.A. ($^\circ$) & -180 & 180 & 1\\
$s$ (mas)  & 0 & 700 & 5\\
$\Delta$ (mag)  & 3 & 7 & 0.05
\enddata
\\
\end{deluxetable}

To search for companions in our data, we fit binary models to our kernel phases. Given the angular resolution of our observations, a binary can be approximated as two delta functions, the Fourier transform of which is an analytic function. Equation (\ref{eq-fft}) gives the complex visibility for a binary with a companion separation $s$, position angle P.A. (measured E of N), and brightness ratio $b$ (in units of the primary's brightness).
\begin{equation}
V(u,v) = \frac{1}{\sqrt{2 \pi}}\left(1 + b~e^{i\cdot s\left(u\cdot sin\left(PA\right) + v\cdot cos\left(PA\right)\right)}\right)
\label{eq-fft}
\end{equation}
The phase measured for a binary by a baseline with coordinates $(u,v)$, is then the angle of Equation \ref{eq-fft}:
\begin{equation}
\phi(u,v) = tan^{-1}\left(\frac{Im(V(u,v))}{Re(V(u,v))}\right).
\label{eq-ph}
\end{equation}

To create a model set of kernel phases, we calculate the closure phases for each triangle using the locations and sizes of the mask sub-apertures, and sky rotation angles included in our observations. We then project the closure phases into kernel phases in the same way as we have done for the data. Due to the symmetry of the Fourier transform, each closure phase can correspond to one of two closing triangles. To keep the sign of our closure phases (and thus kernel phases) consistent between data and model, we sample the same closing triangles in both. Additionally, where necessary, we use observations of a known binary to calibrate the orientation of our detector on the sky. For the NaCo data, as in Huelamo et al. 2011, we use observations of the binary $\theta$ Ori C, taken in 2010 April.

We perform $\chi^2$ fitting of the binary models to each individual dataset using both a grid and nested sampling \citep{sands06}. Our grid spans a range of parameters in binary position angle (P.A.), separation ($s$), and contrast in magnitudes ($\Delta$). $\Delta$ can be related to $b$, the brightness ratio, by the following:
\begin{equation}
\Delta = -2.5~log_{10}(b)
\end{equation}
Table \ref{tab-bgrid} lists the ranges and spacings for each parameter. For each set of parameters, we calculated model kernel phases and a $\chi^2$ statistic. We used the best grid fit as an input for our nested sampling algorithm.

Nested sampling involves first filling the parameter space with a large number (in our case, 100) of points, and calculating a likelihood ($exp(-\chi^2/2$)) for each point. We then replace the lowest likelihood point with a random member of the remaining 99, and evolve it using a Markov-Chain to a higher likelihood region of the parameter space. We repeat this process until the 100 points satisfy a convergence criterion, in our case, the scatter in the ensemble must be a small fraction ($\sim 0.1 \%$) of the mean. We give one of the 100 nested sampling points the best grid fit as an initial value, and assign random values to the remaining 99. This is not necessary for fits in which there is one clear minimum, but it can help in preventing the nested sampling fit from converging to local minima.

We report parameter errors calculated from a $\chi^2$ interval. After finding the minimum $\chi^2$ using nested sampling, we then scale all $\chi^2$ values for a grid of parameters so that the reduced $\chi^2$ of the best-fit model is equal to 1. With the scaled set of $\chi^2$ values, we find all grid points within $\Delta\chi^2$ of 3.53 \citep{numrecipes} to place a $1 \sigma$ error on the fit parameters. 

Bootstrapping often gives the most conservative estimates of parameter errors. However, as noted in \citet{numrecipes}, the results of data fitting in Fourier space rely heavily on all grid points being present, and while the $(u,v)$ coverage in the NaCo and MagAO data sets is good, it is by no means complete. For this reason, bootstrapped data sets do not fairly represent the noise in the data. We confirmed this by bootstrapping Gaussian noise sampled at the same $(u,v)$ points as each dataset; a given noise realization's best fit contrast ratio was much higher than the bootstrapped distribution would suggest.

\renewcommand{\arraystretch}{1.8}
\begin{deluxetable*}{lccccccc}
\tabletypesize{\scriptsize}
\tablecaption{Binary Fit Results}
\tablewidth{0pt}
\tablehead{
\colhead{Data Set} & \colhead{P.A. (deg)} & \colhead{Separation (mas)} & \colhead{$\Delta L'$} & \colhead{P(FA)\tablenotemark{a} (\%)} &  \colhead{P(FA)\tablenotemark{b} (\%)} & \colhead{$\Delta_{\chi^2,Hu}$} & \colhead{$\sigma_{Hu,allowed}$}\\
}
\startdata
\sidehead{\textbf{2010 March 17: VLT/NaCo L$'$}}
$\mathbf{\Delta t^{-1}}$ & \textbf{83} $\mathbf{\pm^{7}_{9}}$ & \textbf{88} $\mathbf{\pm^{22}_{58}}$  & \textbf{5.5} $\mathbf{\pm^{0.5}_{2.5}}$  & \textbf{11} & $\mathbf{<0.1}$ & \textbf{2.63} & \textbf{1}\\
$\Delta t^{-10}$ & 82 $\pm^{5}_{7}$ & 78 $\pm^{22}_{48}$& 5.3 $\pm^{0.6}_{2.3}$ & 4  & $<0.1$ & 2.56 & 1\\
\hline

\sidehead{\textbf{2011 March 14: VLT/NaCo L$'$}}
$\mathbf{\Delta t^{-1}}$ & \textbf{92} $\mathbf{\pm^{11}_{14}}$ & \textbf{87} $\mathbf{\pm^{33}_{57}}$& \textbf{5.3} $\mathbf{\pm^{0.7}_{2.3}}$ & \textbf{25} & \textbf{18} & \textbf{2.54} & \textbf{1}\\
$\Delta t^{-10}$ & 92 $\pm^{117}_{21}$ & 94 $\pm^{356}_{64}$& 5.4 $\pm^{0.8}_{2.4}$ & 18 & 36 & 1.25 & 1\\
\hline

\sidehead{\textbf{2012 March 8: VLT/NaCo L$'$}}
$\mathbf{\Delta t^{-1}}$ & \textbf{-40} $\mathbf{\pm^{5}_{5}}$ & \textbf{32} $\mathbf{\pm^{28}_{2}}$& \textbf{3.4} $\mathbf{\pm^{1.6}_{3.2}}$ & \textbf{1.3} & $\mathbf{<0.1}$ & \textbf{23.10} & $\mathbf{>4}$\\
$\Delta t^{-10}$ & -38 $\pm^{5}_{5}$ & 29 $\pm^{54}_{6}$& 0.4 $\pm^{5.3}_{0.3}$ & 4.3 & $<0.1$ & 20.43 & 4\\
\hline

\sidehead{\textbf{2013 March 25: VLT/NaCo L$'$}}
$\mathbf{\Delta t^{-1}}$ & \textbf{83} $\mathbf{\pm^{3}_{1}}$ & \textbf{55} $\mathbf{\pm^{25}_{25}}$& \textbf{5.2} $\mathbf{\pm^{0.7}_{1.9}}$ & \textbf{0.9} & $\mathbf{<0.1}$ & \textbf{25.82} & $\mathbf{>4}$\\
$\Delta t^{-10}$ & 82 $\pm^{4}_{4}$ & 25 $\pm^{50}_{0.01}$& 2.8 $\pm^{3.1}_{0.2}$ & 3.4 & $<0.1$ & 17.35 & 4\\
\hline

\sidehead{\textbf{2013 March 26: VLT/NaCo Ks}}
$\mathbf{\Delta t^{-1}}$ & \textbf{74} $\mathbf{\pm^{4}_{0.3}}$ & \textbf{42} $\mathbf{\pm^{8}_{22}}$& \textbf{5.2} $\mathbf{\pm^{0.3}_{1.8}}$ & \textbf{1.3} & $\mathbf{<0.1}$ & \textbf{21.47} & \textbf{4}\\
$\Delta t^{-10}$ & 77 $\pm^{1}_{3}$ & 51 $\pm^{9}_{11}$& 5.4 $\pm^{0.2}_{0.3}$ & 0.4 & $<0.1$ & 10.64 & 2\\
\hline

\sidehead{\textbf{2013 April 5: Magellan/MagAO/Clio2 L$'$}}
$\mathbf{\Delta t^{-1}}$ & \textbf{112} $\mathbf{\pm^{176}_{99}}$ & \textbf{337} $\mathbf{\pm^{153}_{147}}$& \textbf{5.8} $\mathbf{\pm^{0.6}_{0.4}}$  & \textbf{32} & \textbf{2} & \textbf{15.97} & \textbf{4} \\
$\Delta t^{-10}$ & -131 $\pm^{3}_{5}$ & 315 $\pm^{25}_{26}$& 5.8 $\pm^{0.5}_{0.3}$  & 32 & $ 4.5 $ & 12.85 & 3
\enddata
\label{tbl-binfits_good}
\tablenotetext{a}{Using distribution of noise simulation best fits}
\tablenotetext{b}{Using distribution of noise simulation $F$ statistics}

\end{deluxetable*}
\renewcommand{\arraystretch}{1.0}

\section{Noise Simulations}\label{sec-sims}

For each dataset, we fit simulated kernel phases to quantify our type I (false-positive) and type II (false-negative) errors. The contrast ratio and separation parameters in a binary model act much like the amplitude and frequency of a sine wave; they take on non-zero values when fit to noise. In order to determine the separations and contrast ratios that could be caused by the noise levels in these data sets, we simulated Gaussian kernel phases from distributions fit to the data. These are shown in Figure \ref{fig-goodcal} for L$'$ and 2013 Ks data, and in Appendix \ref{app-otherdata} for 2010 and 2011 Ks and H band data. We use $(u,v)$ coverage and sky rotation identical to each observation. We fit binary models to 1000 of these noise realizations and create a probability distribution from the best fits. Since the best fit position angle for the noise simulations is uniformly distributed, we create two-dimensional confidence intervals from the best fit separations and contrasts. We then compare the best fit from our data to these confidence intervals. For example, if our best fit lies just outside the contour enclosing 95\% of the best fits, there is a 5\% chance that the fit is drawn from the distribution and thus a 5\% chance that the fit resulted from noise alone. 

A second way of quantifying type I errors is to calculate the F statistic, the best fit $\chi^2$ divided by the null model $\chi^2$, for each noise realization. Comparing the F statistic for a dataset's best fit to a distribution of these simulated noise F statistics yields the probability that the best fit incorrectly rejects the null hypothesis. For example, if 95\% of the simulated noise realizations have lower F statistics than a dataset's best fit, there is a 95\% probability that the data contain no real signal. This procedure is outlined in detail in \citet{protassov02}. 

To estimate our type II errors - the likelihood that the best fit to our data would be a false negative - we simulated observations with noise plus the signal from a companion. We generate 1000 Gaussian noise realizations and fit a binary model to the simulated data. We take the fraction of fits where the input signal was not recovered to be our type II error. We also calculate F statistics from the noise + signal realizations, for comparison with both the noise F statistics and the data's best fit F statistics.

\begin{figure*}[h!]
\epsscale{0.95}
\plotone{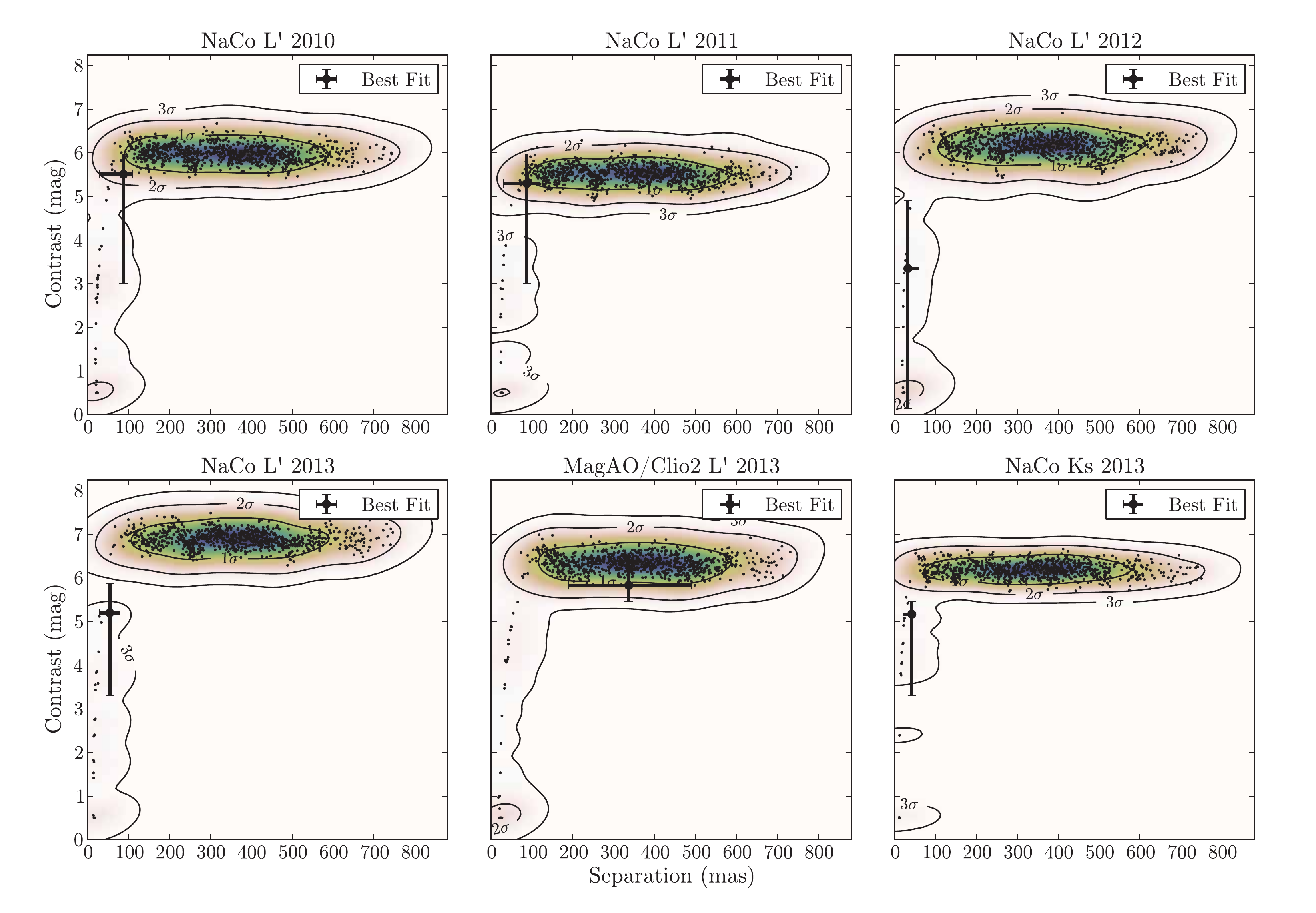}
\caption{Noise simulations for all L$'$ and NaCo 2013 Ks data sets. The scattered points show the best fits to 1000 noise realizations for each dataset, drawn from the Gaussian distributions shown in Figure \ref{fig-goodcal}. The color scale shows the probability distribution interpolated from the best fits, while the contours indicate $1\sigma$, $2\sigma$, and $3\sigma$ confidence intervals. The bold point with error bars represents the best fit for each epoch.}\label{fig-good_n_sims}
\end{figure*}

Figure \ref{fig-good_n_sims} shows the noise simulations for all L$'$ and 2013 Ks data sets with a $\Delta t^{-1}$ calibration. The scattered points show the results of 1000 fits to Gaussian noise comparable to the scatter in each dataset, and the points with error bars show the best fit to each dataset. The color scale shows a probability distribution interpolated from the results of the simulations, and the contour lines indicate 1$\sigma$, 2$\sigma$, and 3$\sigma$ confidence intervals. 

\begin{figure*}[h!]
\epsscale{1.0}
\plotone{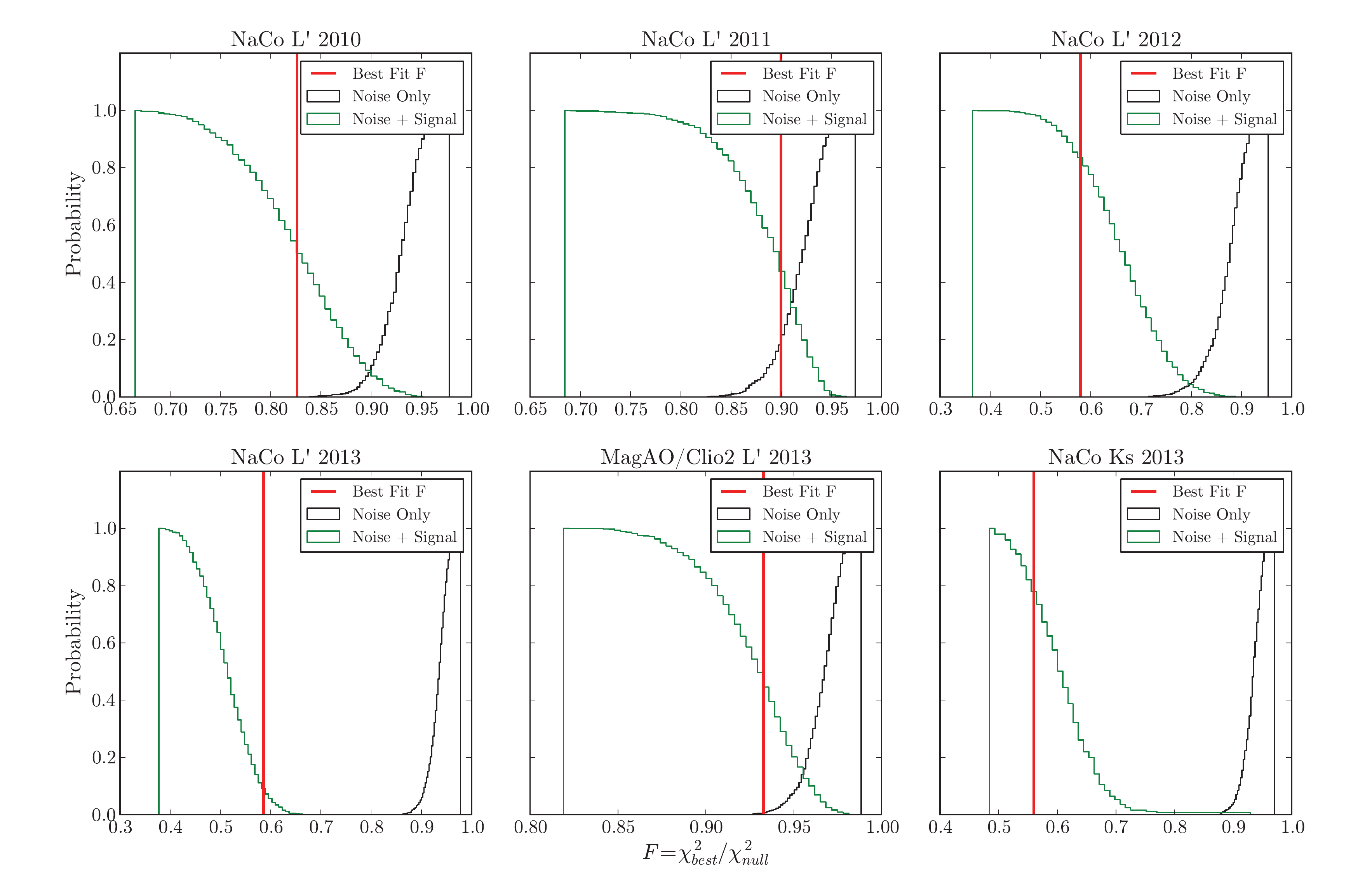}
\caption{False alarm testing for all L$'$ data sets and NaCo 2013 Ks observations. In each panel, the black, cumulative histogram shows the F statistics (best fit $\chi^2$ divided by null model $\chi^2$) for 1000 noise simulations. The intersection of the red, vertical line with the black histogram yields a false alarm probability. The green, cumulative histogram shows the F statistics from the 1000 noise + signal simulations carried out for each set of observations. The intersection of the red line with this histogram gives the fraction of noise + signal simulations that look less significant (higher F statistic) than the best fit.}\label{fig-good_FAs}
\end{figure*}

Figure \ref{fig-good_FAs} shows the false alarm probabilities for the same data sets in Figure \ref{fig-good_n_sims}. In each panel, the black line is a histogram of all of the F statistics for 1000 Gaussian noise realizations. The red, vertical lines indicate the F statistic for each dataset's best fit; the intersection of these lines with the black histograms gives the probability that the best fit to the data resulted from noise. The green, cumulative histogram shows the distribution of F statistics for the 1000 noise + signal realizations. The intersection of the red line with this distribution gives the fraction of noise + signal realizations with fits that look less significant (higher F statistics) than our best fit. For Figures \ref{fig-good_n_sims} and \ref{fig-good_FAs}, the noise properties of the $\Delta t ^{-10}$ calibration yielded comparable results (see Table \ref{tbl-binfits_good}).

While using a $\chi^2$ statistic to assign significance depends on the data error bars, which could be difficult to estimate properly, these simulations take into account only the scatter in the kernel phases. For this reason, in Section 8, we use the results of the noise simulations to estimate the significance of the best fit binary results.

\section{Results}\label{sec-res}

Table \ref{tbl-binfits_good} shows our results in chronological order. The first three columns list the results of binary fits to the kernel phases. Following the binary fits, the next columns list two false-alarm probabilities, the first calculated using the distribution of best fits to noise realizations and the second using the distribution of best fit F statistics (see Section \ref{sec-sims}). The last two columns list the $\Delta \chi^2$ corresponding to the Huelamo et al. (2011) binary for that dataset, alongside the corresponding confidence interval at which it is allowed. The bold values are binary fit results for the $\Delta t^{-1}$ calibrated data. We include the best fits from this calibration strategy in Section 9.

\subsection{2010 VLT/NaCo $L'$ Data - Initial Detection}\label{subsec-degen}

\begin{figure}
\epsscale{1.2}
\plotone{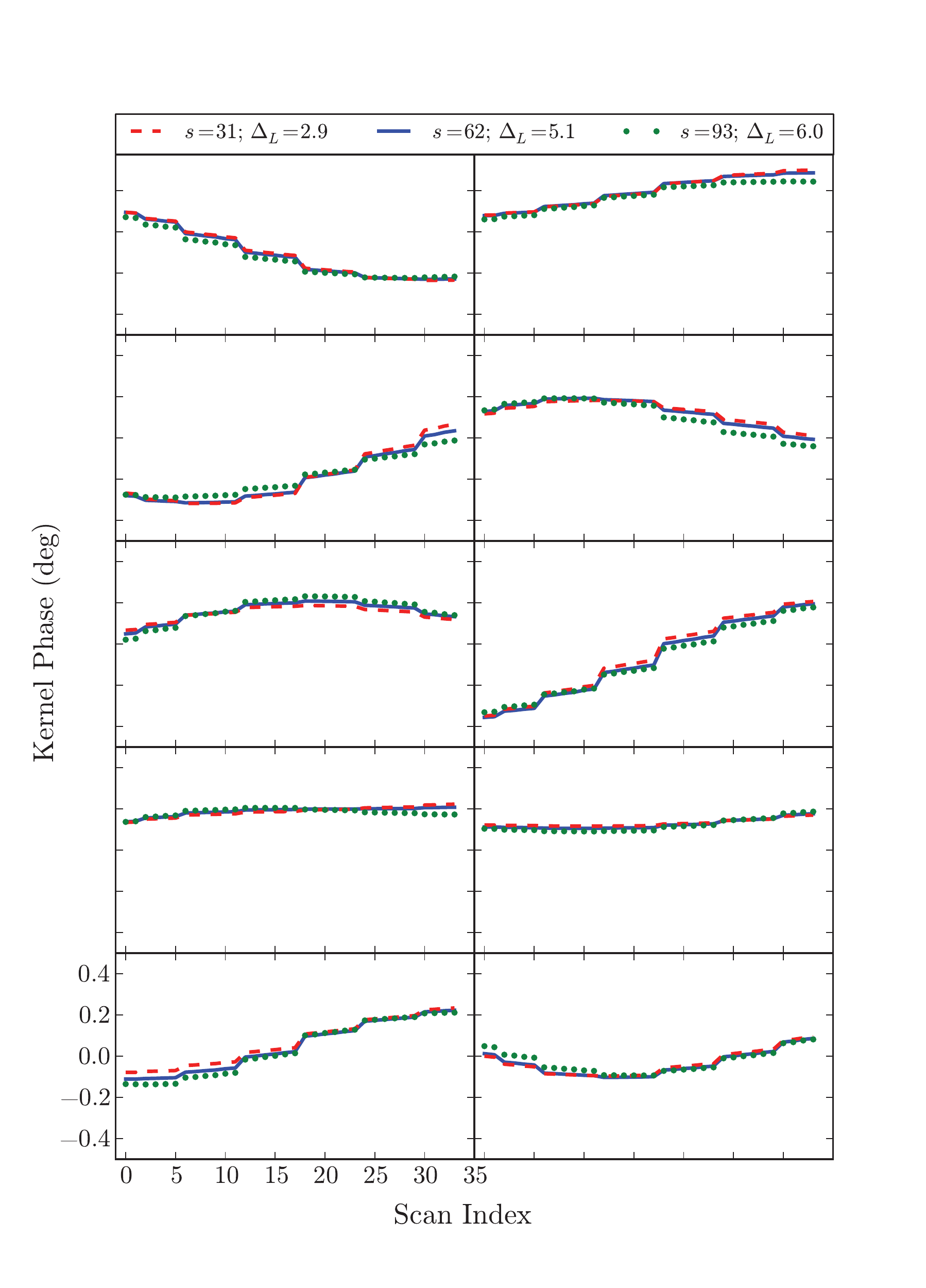}
\caption{Separation - contrast degeneracy. The subplots each show the evolution of one simulated kernel phase during the MagAO/Clio2 observations. The different lines are binaries with identical position angles, but different separations and contrast ratios. The blue solid line shows the signal from a companion similar to that published in \citet{hu11} (s = 62 mas, $\Delta L' = 5.1$), the red dashed line shows s = 31 mas and $\Delta L' = 2.9$, and the green dotted line shows s = 93 mas and $\Delta L' = 6.0$. While slight differences exist between the binary models, they are not detectable given the accuracy of the data.}.
\label{fig-degen}
\end{figure}

For these data, we first fit our neighbor-calibrated (weighting by $\Delta t^{-10}$) kernel phases, since the calibration strategy for the closure phases in \citet{hu11} is most similar to this approach \citep[see also][]{lacour11}. We find a best fit binary with position angle of $82\pm~^{5}_{7}$$^\circ$, separation of $78\pm~^{22}_{48}$ mas, and $\Delta L' = 5.3\pm~^{0.6}_{2.3}$ mag. This is comparable to the Huelamo et al. (2011) best fit - position angle of 78$\pm1^\circ$, separation of 62$\pm7$ mas (6.7 AU at 108 pc), and $\Delta L'$ of 5.1$\pm~0.2$ mag. Our fit parameter errors are substantially larger than those quoted in \citet{hu11}, which were derived using $\chi^2$ intervals from a binary fit to closure phases \citep[see also][]{lacour11}. Closure phases have correlated errors, which could bias the parameter errors derived using a simple $\chi^2$ fit. Parameter errors also depend on whether the data error bars have been scaled such that the reduced $\chi^2$ is equal to 1, which is appropriate if you assume the model to be correct, and that underestimated error bars are causing a high reduced $\chi^2$. Our large separation and contrast error bars are consistent with the severe degeneracy between these two parameters. Figure \ref{fig-degen} illustrates this degeneracy, with kernel phases plotted for separations $0.5 \times$, $1\times$, and $1.5\times$ the separation of the \citet{hu11} binary, varying the contrast ratio. The three models are nearly indistinguishable. For this reason, we believe our fit errors to be a more realistic representation of the parameter constraints.

After the nearest-neighbor fit, we then fit the $\Delta t^{-1}$ calibrated kernel phases, finding a best fit with position angle of $83\pm~^{7}_{9}$$^\circ$, separation of $88\pm^{22}_{58}$ mas, and $\Delta L' = 5.5\pm^{0.5}_{2.5}$ mag. The $\Delta t^{-1}$ calibration method resulted in kernel phases with lower scatter than the $\Delta t^{-10}$ method - $0.63^\circ$ versus $0.66^\circ$. Using a $\chi^2$ interval, the Huelamo et al. (2011) binary model is within $1 \sigma$ of the best fit for both calibration methods. Appendix \ref{app-data_mod} shows our $\Delta t^{-1}$ calibrated kernel phases with both the Huelamo et al. (2011) model and the best fit model from this work. The two models are nearly indistinguishable.

Figure \ref{fig-good_n_sims} shows that there is a small probability that the binary detection resulted from a random noise fluctuation. The best fit falls on a contour that encloses 89\% of the simulation results; there is an 11\% chance that the 2010 fit was the result of noise. The 2010 L$'$ best fit is then significant at roughly the $2\sigma$ level. The distribution of F statistics, shown in Figure \ref{fig-good_FAs}, indicates that the false alarm probability for this dataset is $<0.1\%$, giving the fit $\sim 3\sigma$ significance.

\subsection{VLT/NaCo 2011 L$'$ Data}

For these data, the $\Delta t^{-1}$ best fit has a position angle of $92\pm^{11}_{14}$$^\circ$, a separation of $87\pm^{33}_{57}$ mas, and a contrast of $\Delta L' = 5.3\pm^{0.7}_{2.3}$ mag. Compared to the $\Delta t^{-10}$ calibration (listed in Table \ref{tbl-binfits_good}), the $\Delta t^{-1}$ weighting reduced the number of outliers in the calibrated kernel phases, providing a tighter constraint on the binary fit parameters. Again, for both of these calibration methods, the Huelamo et al. (2011) model is within $1 \sigma$ of the best fit. Appendix \ref{app-data_mod} shows our calibrated kernel phases with both the Huelamo et al. (2011) model and the best fit model from this work. 

Figure \ref{fig-good_n_sims} shows the results of the noise simulations for this dataset. The distribution of best fits suggests that the fit to the data is significant at the $1\sigma$ level; the point with error bars falls on a contour which encloses 75\% of the fits to noise, giving a 25\% false alarm probability. This agrees roughly with the F statistic distribution, shown in Figure \ref{fig-good_FAs}. The best fit F statistic, $F = 0.900$, gives an 18\% probability of false alarm. This corresponds to a $1-2\sigma$ confidence level. Here, the best fit is consistent with the distribution of noise + signal F statistics, which overlaps with the noise-only F statistics. The overlap of these two distributions indicates that, with the properties of the NaCo 2011 data, noise alone can produce best fits that appear as significant as noise plus the Huelamo binary model.

\subsection{VLT/NaCo 2012 L$'$ Data}

The $\Delta t^{-1}$ fit resulted in a position angle of $-40\pm^{5}_{5}$$^\circ$, separation of $32\pm^{28}_{2}$ mas, and a contrast of $3.4\pm^{1.6}_{3.2}$ mag. The kernel phases, with this work's best fit and the Huelamo et al. (2011) companion signal over-plotted, are shown in Appendix \ref{app-data_mod}. Using a $\chi^2$ interval, these data rule out the presence of the Huelamo signal at greater than 4$\sigma$.

The results of noise simulations for the 2012 L$'$ data are shown in Figure \ref{fig-good_n_sims}. The best fit lies on a contour that encloses 98.7\% of the fits to noise, giving a 1.3\% false alarm probability. This suggests that the best fit is significant at nearly $3\sigma$. The F statistics (see Figure \ref{fig-good_FAs}) also indicate that the fit is $\sim3\sigma$ significant; less than 0.1\% of F statistics are lower than that for the best fit ($F = 0.579$). 

\subsection{VLT/NaCo 2013 L$'$ Data}

The $\Delta t^{-1}$ best fit has a position angle of 83$\pm^{3}_{1}$$^\circ$, separation of 55$\pm^{25}_{25}$ mas, and a contrast of 5.2$\pm^{0.7}_{1.9}$ mag. Appendix \ref{app-data_mod} shows the kernel phases with our best fit model (red line) plotted alongside the Huelamo et al. (2011) binary model. The Huelamo et al. (2011) model is ruled out at $\sim4$ sigma using a $\chi^2$ interval; the separation and contrast are within $1\sigma$ of the Huelamo et al. (2011) binary, but the position angle is greater by 5.46$^\circ$. However, adopting our larger parameter error bars, (see $\S$8.1), the two models are consistent at $1\sigma$. 

Our best fit to these data lies on a contour enclosing 99.1\% of the points, giving a 0.9\% chance that the best fit resulted from noise alone (see Figure \ref{fig-good_n_sims}). This suggests that the best fit is significant at nearly $3\sigma$. The F statistics, shown in Figure \ref{fig-good_FAs}) give a lower false alarm probability than this. The best fit F statistic (F = 0.586) is lower than all 1000 noise simulation F statistics, giving a $<0.1\%$ false alarm probability. 

\subsection{VLT/NaCo 2013 Ks Data}

The $\Delta t^{-1}$ best fit parameters are position angle of 74.45$\pm^{3.72}_{0.31}$$^\circ$, separation of 42.07$\pm^{7.93}_{22.07}$ mas, and contrast of 5.17$\pm^{0.29}_{1.87}$ mag. These data (scattered points) with our best fit model and the Huelamo et al. (2011) model are shown in Appendix \ref{app-data_mod}. The error bars for this work's best fit and the Huelamo et al. (2011) best fit overlap for both position angle and contrast. However, the best fit separation is smaller than the Huelamo et al. (2011) model. Given the large error bars, this difference is significant at less than 2$\sigma$.

The best fit to these data falls on a contour that encloses 98.7\% of the noise simulation results (see Figure \ref{fig-good_n_sims}). This gives a 1.3\% chance that the best fit was the result of noise. The best fit F statistic ($F = 0.560$) was lower than all of the simulated F statistics, shown in Figure \ref{fig-good_FAs}. This gives a $<0.1\%$ probability of the best fit resulting from noise. These estimates suggest that the fit to these data is significant at the $\sim3\sigma$ level.

\subsection{Magellan/MagAO/Clio2 2013 Data}

The $\Delta t^{-1}$ calibration for these data has a best fit with a position angle of $112\pm^{176}_{99}$$^\circ$, separation of $337\pm^{153}_{147}$ mas, and $\Delta L' = 5.8\pm^{0.6}_{0.4}$ mag. Both this and the $\Delta t^{-10}$ best fit (see Table \ref{tbl-binfits_good}) are unreasonable in that binaries with these parameters should have been, but were not detected in the 2013 MagAO/Clio2 direct imaging data. 

Shown in Figure \ref{fig-good_n_sims}, the 2013 MagAO/Clio2 best fit lies on a contour that encloses 68\% of the simulated fits, giving a 32\% probability that it could have resulted from noise - a $1\sigma$ result. For these data, this method does not agree well with the F statistic false-alarm estimation. The best fit F statistic, shown in Figure \ref{fig-good_FAs}, is greater than only 2\% of the simulated F statistics, suggesting that there is a 2\% probability that the best fit was caused by noise. We speculate that this discrepancy could be caused by outliers in the data themselves. The null model $\chi^2$ for a set of kernel phases with non-Gaussian outliers will be greater than the null model for kernel phases drawn from a Gaussian distribution, which could reduce the F statistic.  Figure \ref{fig-goodcal} shows that the MagAO data could indeed have both outliers and a small non-zero mean kernel phase. These could both inflate the null model $\chi^2$ compared to that for Gaussian data, making the F statistic method less reliable as a false alarm probability estimator. 

Appendix \ref{app-data_mod} shows our calibrated kernel phases with the NaCo 2013 L$'$ best fit and the best fit model. The NaCo 2013 L$'$ best fit is allowed at 3$\sigma$ by a secondary $\chi^2$ minimum MagAO/Clio2 data. We estimated our type II errors following Section \ref{sec-sims}, adding the 2013 NaCo L$'$ best fit to 1000 noise realizations. The results show that the probability of missing this signal, had it been present in these data, is 49.9\%. Thus, the 2013 NaCo L$'$ best fit is allowed by the MagAO/Clio2 observations.

\section{Discussion}\label{sec-disc}

The NaCo 2012 and MagAO/Clio2 2013 L$'$ data sets have best fit binaries that are inconsistent with both the Huelamo et al. (2011) model and our best fit to the NaCo 2013 L$'$ data. However, simulations show that there is a non-zero chance that these fits resulted from noise - 1.3\% for the NaCo and 32\% for the MagAO/Clio2 observations. We simulated noise realizations (see Section \ref{sec-sims}) to estimate our type II errors for both of these data sets, using the NaCo 2013 L$'$ signal as the input binary model. The chance that we would have missed the binary signal in these data is 15.8\% for NaCo 2012 and 49.9\% for MagAO/Clio2 2013.

Keeping these type II errors in mind, in the subsections concerning orbital motion and forward scattering we first assume that the tentative NaCo 2012 detection is reliable but that the MagAO/Clio2 non-detection is due to noise. We then discuss the results assuming that noise fluctuations led to a false detection in the NaCo 2012 data, while a signal compatible with the other NaCo data sets was actually present beneath the noise. We also consider the possibility that asymmetries caused by planet-disk interactions could have caused the observed kernel phases. In the last subsections, we discuss whether a chance alignment or systematic error could masquerade as a companion in the data sets where we detect a significant signal. 

\subsection{Companion Orbital Motion}

\begin{figure*}
\epsscale{1.2}
\plotone{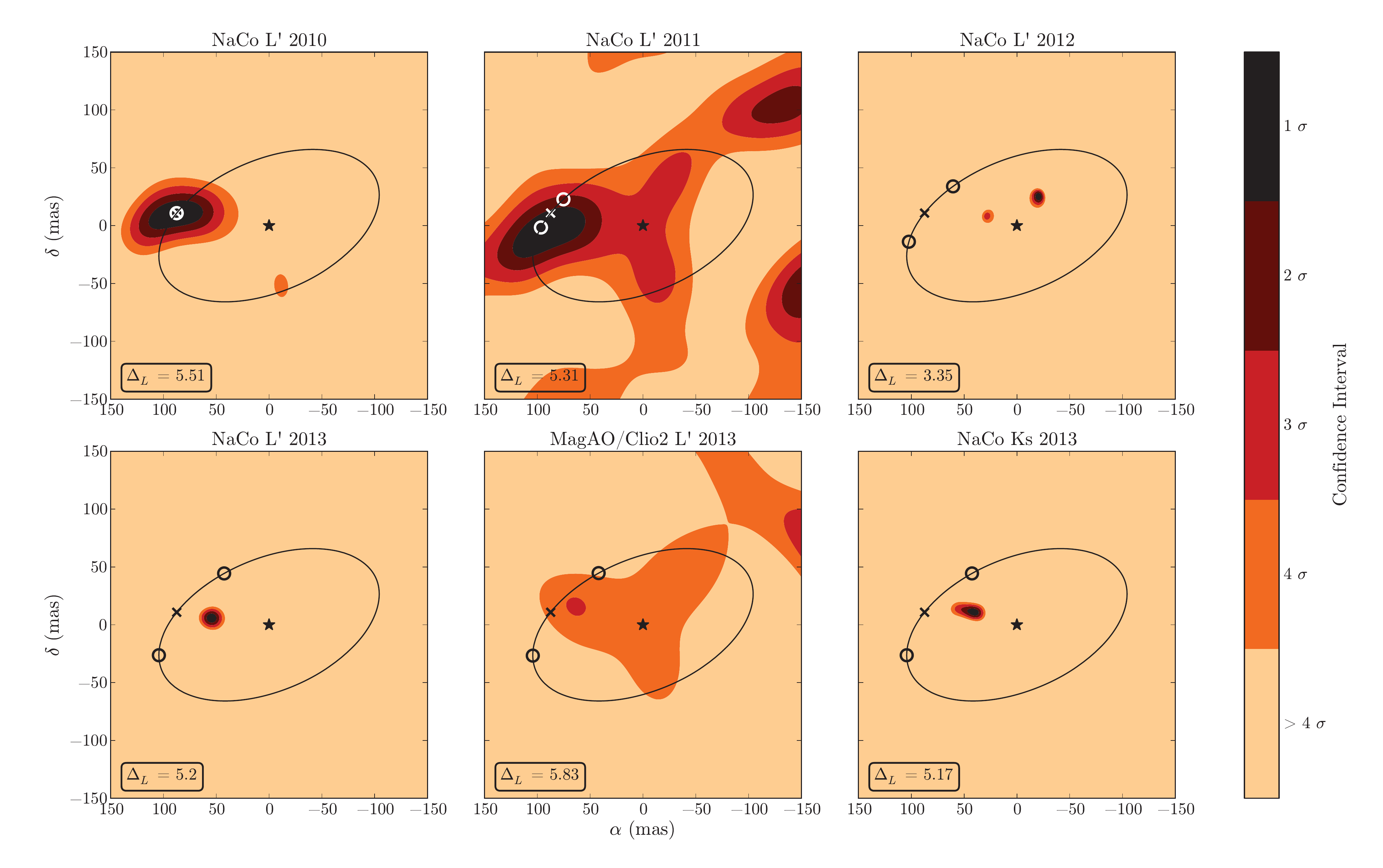}
\caption{$\chi^2$ slices at the fixed contrast ratio for all L$'$ observations as well as 2013 NaCo Ks observations, with filled contours at $1$ to $ >4\sigma$ confidence limits. The line indicates a circular orbit in the plane of the outer disk \citep[see][]{olofsson13}. The $\times$s show the initial position of the putative companion from our re-reduction of the 2010 NaCo L$'$ data, while the circles show the predicted position(s) of a planet on the orbit. We plot two since the planet could be orbiting in either direction.}
\label{fig-good_slices}
\end{figure*}

\begin{figure}[h!]
\epsscale{1.25}
\plotone{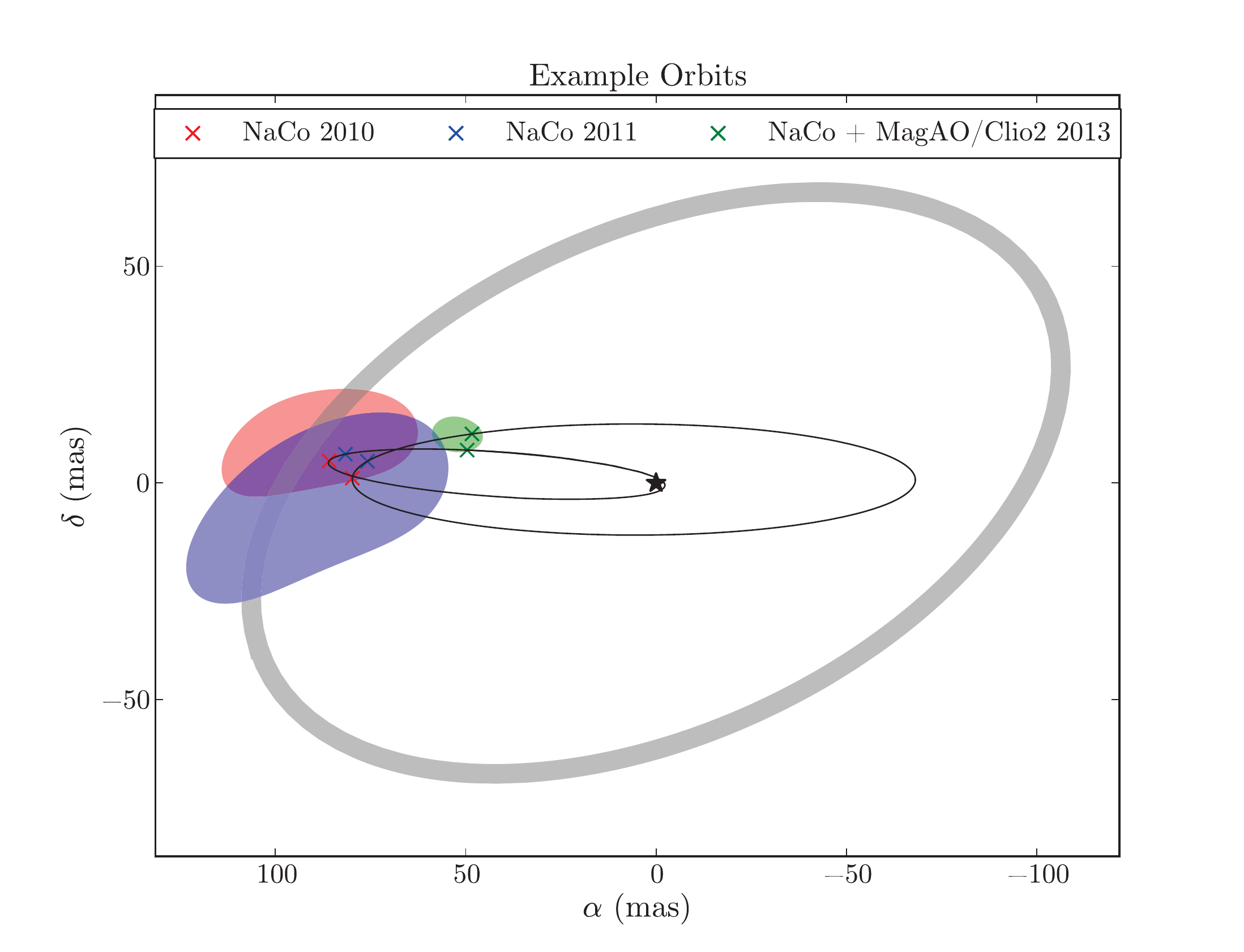}
\caption{Example orbits consistent with 2010, 2011, and 2013 NaCo best fits. The shaded regions show the $1\sigma$ confidence intervals at the best-fit contrast ratio for the 2010 NaCo L$'$ (red), 2011 NaCo L$'$ (blue), and 2013 NaCo L$'$+Ks (green) data sets. Colored points mark the predicted orbital positions at the times of the observations. Black ellipses mark the full orbits. The thick, gray line indicates the inner edge of the outer disk, as determined by Olofsson et al. (2013).}
\label{fig-orbits}
\end{figure}

Detecting orbital motion of a binary signal in multi-epoch data sets would confirm the claimed companion from Huelamo et al. (2011). For example, \citet{ki12} detected a planet candidate in the LkCa15 transition disk. Multi-epoch observations of this object revealed orbital motion of the companion at the level of $\sim4^\circ$ a year \citep{ik14}, strengthening the case for the protoplanet explanation of the phase signal.  

We compare the predicted position of an orbiting companion to our fit results. Our best fit to the archival discovery data has a separation of 88 mas, which at T Cha's distance of 108 $\pm$ 9 pc \citep{torres08} corresponds to 9.5 AU. We can use the orientation of the outer disk as well as the best-fit separation to predict the position of the planet in followup data sets. 

We first take a circular orbit at the same inclination ($i=58^\circ$) and position angle ($-70^\circ$) as the outer disk as determined by \citet{olofsson13}. We choose the 2010 companion location as the initial orbital position, and then predict the projected separations for the followup data sets. Each panel in Figure \ref{fig-good_slices} shows the orbital prediction for a single observation over a $\chi^2$ slice at its best fit contrast ratio. While the predicted position is within $1\sigma$ of the NaCo 2011 L$'$ best fit, all other data sets, even those with lower false alarm probabilities such as 2013 NaCo L$'$ and Ks, rule out the presence of a companion on this orbit. Thus, this scenario cannot explain the observations.

If we assume the NaCo 2012 L$'$ signal is a real detection, then we need to match its fitted position along with the fits from other epochs. This would require at least one full orbit to be completed between 2010 and 2013. However, the observed projected separations ($s$) at each epoch place a lower bound on the apocenter distance:
\begin{equation}
a(1+e)\geq s.
\label{eq-sbound}
\end{equation}
Using this constraint and the fact that T Cha has a mass of $1.5 M_{\odot}$, we can check whether any orbits with a period of $\sim 3$ yr could produce the observed separations. Even as $e$ approaches 1, an orbit around a 1.5M$_{\odot}$ star cannot have a period of less than $8.5$ yr. This rules out orbital motion as the cause of the different position angle in the 2012 L$'$ data. 

If we assume that the 2012 L$'$ NaCo data missed the signal found in the other data sets, we can ask what orbits would cause the best fits to the remaining observations, which show motion compared to the Huelamo et al. (2011) model. For a grid of orbits, we calculated the predicted positions for the times of the 2010, 2011, and 2013 observations. We compared these positions to the 1$\sigma$ confidence intervals in position angle and separation from the binary fitting. We take the results of a simultaneous fit to the 2013 NaCo and MagAO data as our constraint on the 2013 position. We find that inclined orbits, some of which cross into the outer disk, can produce the observations. Orbits with the same inclination as the outer disk require eccentricities higher than 0.9 to reproduce the observations. Figure \ref{fig-orbits} shows two example orbits over the $1\sigma$ confidence regions at the best fit contrast for the three epochs. While a low-eccentricity orbit in the plane of the disk is inconsistent with the observations, an orbit that is highly eccentric or substantially misaligned with the disk is compatible with the data. Since one would expect a young planet that is still accreting from the disk to be on a low-eccentricity, aligned orbit, these models do not seem physically likely. 

\subsection{Forward Scattering from the Disk}

\citet{olofsson13} found that scattering by dust in the upper layers of T Cha's outer disk could fit the observed phase signal nearly as well as a companion. Indeed, observations of another transition disk, FL Cha, showed that low mass companion and disk-scattering models could both explain the closure phase signal in this single epoch observation \citep{cieza13}. In contrast to the companion model, a constant scattering model would lead to a signal that does not vary in time.

The NaCo 2011 and 2013 data support the hypothesis that a constant level of forward scattering could be responsible for the observations. The error bars for the best fits to these data sets are large enough that they overlap with those for our fit to the NaCo 2010 data. However, the best fit to the 2012 NaCo data is inconsistent with the NaCo 2010 best fit. If we assume that the 2012 NaCo L$'$ detection is not caused by noise, then variability in the amount of scattered light would be required to explain the results. 

The intensity of scattered light in a protoplanetary disk is proportional to the luminosity of the star \citep[e.g.,][]{dong12,inoue08,dandn03}. Stellar variability would lead to variability in the intensity of the scattered light from the disk. However, this would not cause the $\Delta L'$ of a best fit companion to increase, since the ratio of the scattered light intensity to the stellar luminosity would remain constant. Additionally, while T Cha is known to be quite variable in V band, due to changing extinction by circumstellar material, analysis of \emph{Spitzer} spectra along with mid- and far-IR photometry by \citet{schisano09} indicates that it is not variable in the infrared. 

Changing the size distributions, cross sections, or mass fractions of the various dust grain species could change the amount of forward scattering relative to the stellar brightness \citep[e.g.,][]{dalessio98,pollack85}. Changes to the geometry of the disk, such as the height of the outer disk's inner wall, could also alter the scattering intensity \citep[e.g.,][]{dand04}. Some young stars exhibit variability in scattered light due to geometric changes in the inner disk that then shadow the outer parts of the disk \citep[e.g.,]{wisnie08,sitko08,bans12}. These changes can occur on the timescale of weeks to months (the dynamical timescale of the inner disk). However, geometric changes at the radius of the outer disk would take place on the timescale of multiple years. The viscous timescale, an estimate of the time it takes for disk material to shear out, at 9.5 AU, is greater than 100 yr for reasonable values of the viscosity parameter \citep{sands73} and scale-height to radius ratio. The dynamical timescale ($\Omega_k^{-1}$) at the inner edge of the outer disk (12 AU) is 5.41 yr. Thus, disk geometry changes at the radius of the outer disk cannot explain the 2012 NaCo L$'$ best fit.

Scattered light from the upper layers of the outer disk should be brighter at Ks than at L$'$. This is due to the fact that larger grains will settle toward the disk mid-plane and thus contribute less to the total amount of scattered light than smaller grains \citep[e.g.,][]{natta06}. Assuming that the detections in the NaCo 2012 and MagAO data are due to noise, and that forward scattering is causing the other observed signals, one may naively expect the 2013 kernel phase signal to be greater at Ks than at L$'$.  In contrast, we observe similar signal amplitudes at both wavelengths.  However the scattered light may arise from an extended region, significantly resolved by our observations, in which case such a simple interpretation may not apply.  To test this, we simulated scattered light images of T Cha's outer disk using the radiative transfer code Hyperion \citep{robitaille11}. We produced images comparable to those published in Olofsson et al. (2013), using the same dust properties and disk parameters as their best fit. For this model, the mean kernel phase signal was 0.29$^\circ$ at Ks and 0.27$^\circ$ at L$'$. Thus, having similar kernel phase signals for Ks and L$'$ cannot rule out the scattering scenario. We also calculated $\chi^2$ values using the 2013 NaCo observations, and find that the reduced $\chi^2$ for the scattering model is $\sim5.1$. The binary model gives a better fit to the data, but forward scattering can produce kernel phases similar to the observations.

\subsection{Optically Thin Disk Asymmetries}
While the scattering scenario leads to asymmetries in the outer disk, resulting in non-zero kernel phases without the presence of a companion, hydrodynamic simulations \citep[e.g.,][]{fouchet10} suggest that disk-planet interactions can cause asymmetric structures in the optically thin dust within transition disk cavities. Observations in both the infrared \citep[e.g.,][]{muto12} and the sub millimeter \citep[e.g.,][]{isella13,perez14} have confirmed the presence of such asymmetries. Furthermore, recent NRM observations of the transition disk V1247 Orionis revealed phase signals whose best fit binary parameters changed significantly with wavelength. This indicated that the underlying structure was not a simple companion, but asymmetric optically thin material within the disk gap \citep{kraus13}. Image reconstruction would allow us to look for asymmetries in T Cha's disk gap.  We will present image reconstruction in a systematic way in a future paper, but a preliminary effort (using MACIM; \citet{macim06}) suggests a trefoil structure within the cleared region of the transition disk, perhaps compatible with disk-planet interaction models. 

\subsection{Chance Alignment}

We investigate the probability that the chance alignment of a foreground or background object would cause a companion signal in the 2010 NaCo dataset, and that its proper motion would be mistaken for orbital motion between 2010 and 2013. Since T Cha has a proper motion of -39.61 mas yr$^{-1}$ in right ascension and -9.87 mas yr$^{-1}$ in declination, stationary objects in the foreground or background could appear to move like orbiting companions.

The number of chance alignments in a field of view with area $A_{FOV}$ is
\begin{equation}
n_{align} = A_{FOV} \Sigma,
\end{equation}
where $\Sigma$ is the surface density of stars along the line of sight. We use the extent of T Cha's disk gap to define our field of view. From Olofsson et al. 2013, the gap extends from 0.17 AU to 12 AU, subtending $\sim0.11$" at 108 pc. To estimate the surface density of stars along the line of sight, we queried Two Micron All Sky Survey for all stars within 1 degree of T Cha, with $\Delta$Ks lower than 7 (Ks of $\sim 7 - 14$ mag). This gives a stellar surface density of 11544 stars deg$^{-2}$. The probability that a chance alignment with a foreground or background object would cause a companion signal in the NaCo 2010 data is then $\sim 7.0 \times 10^{-6}$. 

\subsection{Systematic Errors}
The 2010, 2011, and 2013 best fits have a low, 0.003$\%$, probability of all resulting from noise, assuming the observations are independent. In this section, we discuss the possibility that the signals could be caused by some systematic effect. 

Systematic errors could be possible since T Cha is such a southern target ($\delta \sim -79^\circ$), and thus transits at 35.27$^\circ$ at the VLT ($\phi = -24.63^\circ$) and slightly higher, 39.90$^\circ$ at Magellan ($\phi = -29.26^\circ$). While our calibrator tests (see Section \ref{sec-calib}) showed that it is unlikely that the calibrators injected a signal into the T Cha kernel phases, this does not rule out all AO-related systematic effects. If we assume a reliable non-detection in the 2013 MagAO/Clio2 data, this could suggest that a systematic error caused the same signal to be present in all of the NaCo data sets. This could be due to T Cha's lower transit elevation as observed from the VLT.

The similar position angles, but different separations of the L$'$ and Ks band best fits (see Figure \ref{fig-good_slices}) also suggest that systematic errors could be an issue. The best fit separation for the 2013 NaCo Ks observations is $\sim42$ mas, while the L$'$ best fit separation is $\sim55$ mas.  The size of an observed speckle pattern is proportional to the wavelength \citep[e.g.,][]{marois00}, and thus we would expect the ratio of a speckle's position in Ks to its position in L$'$ to be $\sim0.57$. The ratio of the NaCo 2013 Ks and L$'$ separations is 0.77. However, the large separation error bars prevent us from placing tight constraints on the relative separations.

\section{Conclusions}\label{sec-conc}
We presented multi-epoch observations of the T Cha transition disk, taken using VLT/NaCo and Magellan/MagAO/Clio2, in L$'$, Ks, and H bands. Out of the nine data sets, three are too noisy to detect signals at the level of the companion candidate published in Huelamo et al. (2011); these are the 2010 and 2011 NaCo Ks and 2013 NaCo H band data. 

We find companion parameters comparable to those published in Huelamo et al. (2011) in our re-reduction of the initial discovery data. Furthermore, we find comparable binary parameters to the companion candidate's in the 2011 NaCo L$'$ dataset, and tentative evidence for radial motion by the time of the 2013 NaCo and MagAO/Clio2 observations. Assuming that noise in the 2012 NaCo and 2013 MagAO/Clio2 data led to non-detections of the signal (simulations show such false non-detections to be fairly probably in these data sets), highly eccentric or misaligned orbits could result in a signal consistent with the observations.

Scattered light from T Cha's outer disk could provide another explanation for the observed kernel phases, although a binary model gives a slightly better fit to the data. Preliminary image reconstructions also suggest an asymmetric structure, perhaps consistent with disk-planet interactions, as the source of the observed signals. Lastly, the ratio of the NaCo 2013 Ks and L$'$ best fit separations argues for the possibility that challenges associated with AO correcting a dim, southern source, could cause a systematic error that would masquerade as a close in companion. The detection of a secondary minimum in the MagAO/Clio2 data at the same position as the NaCo detections is encouraging, but follow-up observations with higher signal to noise are required to rule out the possibility of systematic errors.

\acknowledgments

This work was supported by NSF AAG grant $\#$1211329. J.R.M and K.M.M. were supported under contract with the California Institute of Technology (Caltech) funded by NASA through the Sagan Fellowship Program. This material is based upon work supported by the National Science Foundation Graduate Research Fellowship under Grant No. DGE-1143953. Any opinion, findings, and conclusions or recommendations expressed in this material are those of the authors(s) and do not necessarily reflect the views of the National Science Foundation.

\appendix

\section{Additional Ks and H Band Data Sets}\label{app-otherdata}

\begin{figure*}
\epsscale{1.2}
\plotone{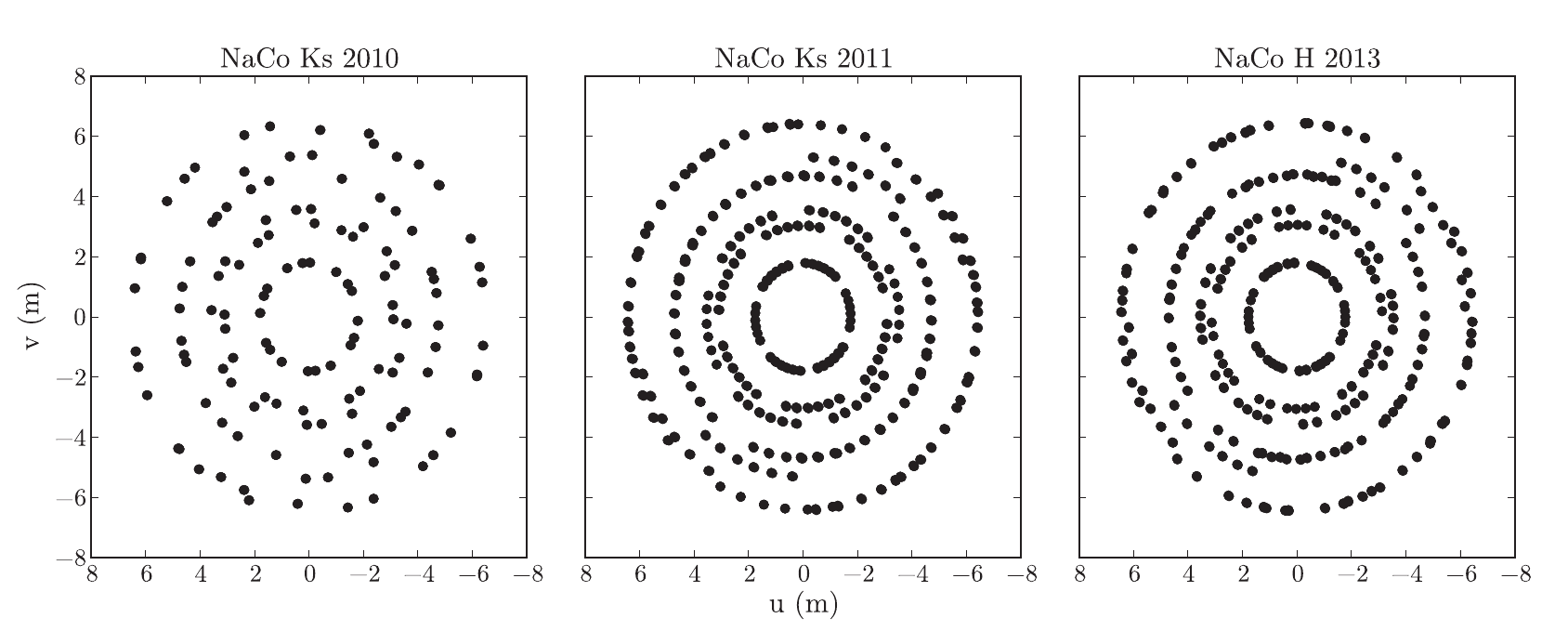}
\caption{Sky rotation comparison in $(u,v)$ space for 2010 and 2011 NaCo Ks data as well as 2013 NaCo H data. The NaCo 2010 and 2011 Ks data had changes in sky rotation of $\sim19^\circ$ and $\sim38^\circ$, respectively. The sky rotation in the NaCo 2013 H data was $\sim39^\circ$.}.\label{fig-uvsL}
\label{fig-bad_rot}
\end{figure*}

\begin{figure}
\epsscale{0.5}
\plotone{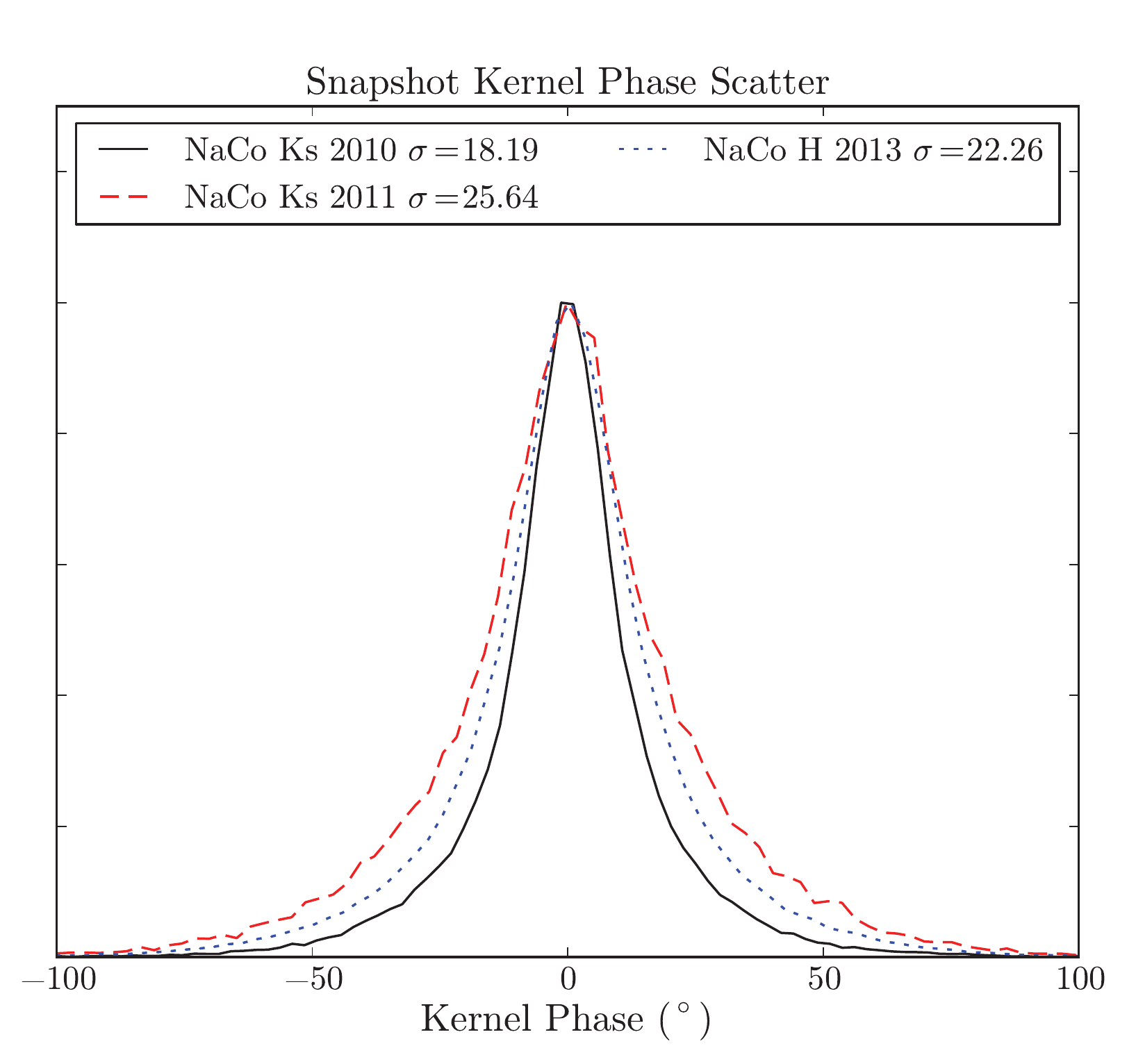}
\caption{Normalized histogram of uncalibrated kernel phases for 2010 and 2011 Ks as well as 2013 H band data. For a subset of each dither (taken so that equal amounts of integration came from each observation), we subtract the mean kernel phase from each individual measurement to generate the histogram shown above. The snapshot errors for these data sets are significantly larger than those for the other six.}\label{fig-bad_snapshot}
\end{figure}

\begin{figure}
\epsscale{0.8}
\plotone{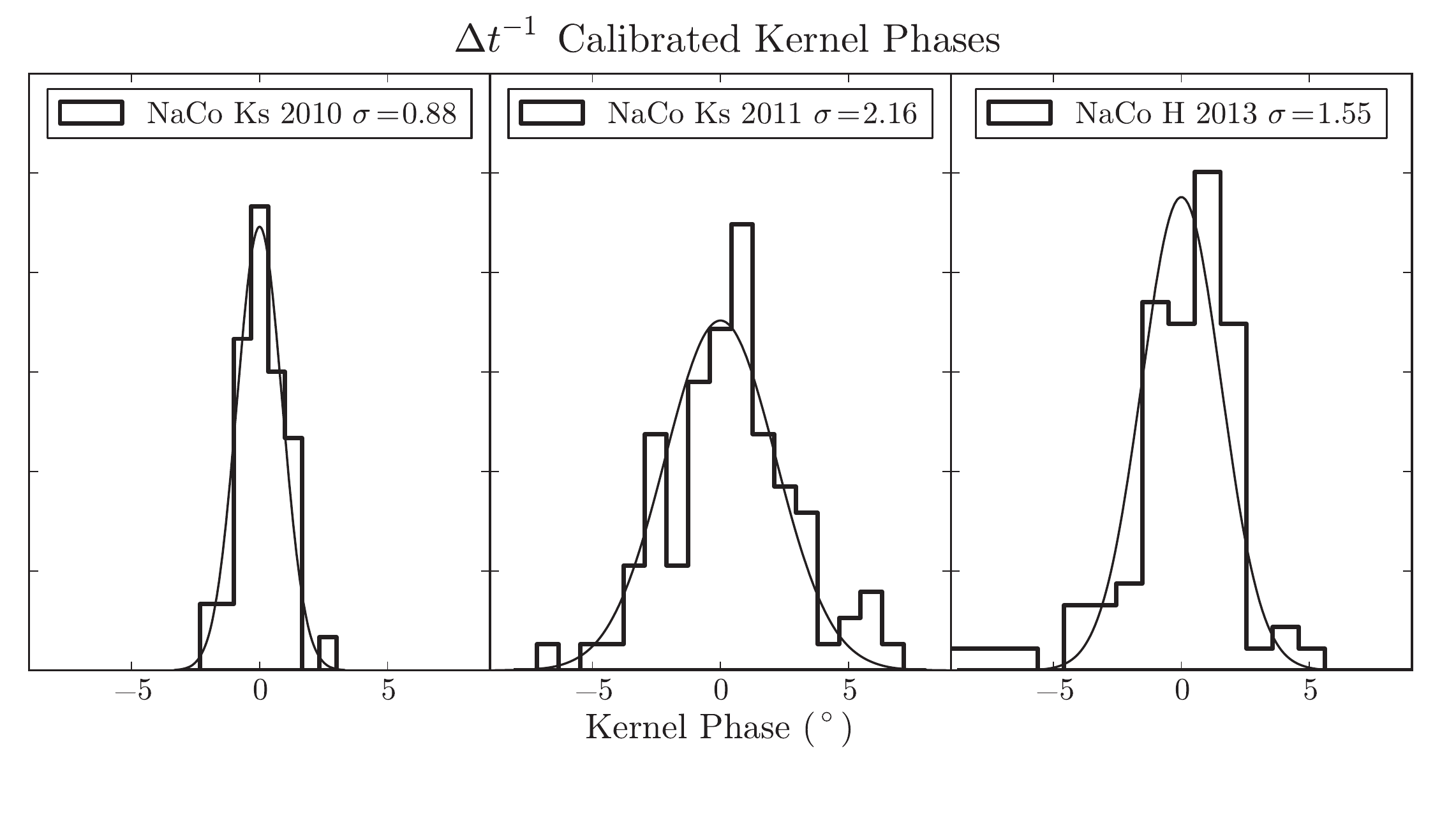}
\caption{Normalized histograms of calibrated NaCo 2010 and 2011 Ks as well as NaCo 2013 H kernel phases, with their best fit Gaussian distributions over plotted. These calibrated kernel phases have much higher scatter than the other data sets. The Gaussian distributions were used to generate noise realizations for the simulations described in Section \ref{sec-sims}.}.\label{fig-badcal}
\end{figure}

Figure \ref{fig-bad_rot} shows the sky rotation coverage for the 2010 and 2011 NaCo Ks and 2013 NaCo H band data sets. The comparison of the snapshot errors (see $\S$\ref{sec-red}) is displayed in Figure \ref{fig-bad_snapshot}. The calibrated kernel phase histograms, with the Gaussian distributions used to generate the noise simulations discussed in Section \ref{sec-sims}, are shown in Figure \ref{fig-badcal}.

\renewcommand{\arraystretch}{1.8}
\begin{deluxetable*}{lcccccccc}
\tabletypesize{\scriptsize}
\tablecaption{Binary Fit Results}
\tablewidth{0pt}
\tablehead{
\colhead{Dataset} & \colhead{PA (deg)} & \colhead{Separation (mas)} & \colhead{$\Delta L'$} & \colhead{P(FA)\tablenotemark{1} (\%)} &  \colhead{P(FA)\tablenotemark{2} (\%)} & \colhead{P(miss) (\%)} & \colhead{$\Delta_{\chi^2,Hu}$} & \colhead{$\sigma_{Hu,allowed}$}\\
}
\startdata
\sidehead{\textbf{2010 July 1: VLT/NaCo Ks}}
$\mathbf{\Delta t^{-1}}$ & \textbf{122} $\mathbf{\pm^{66}_{137}}$ & \textbf{268} $\mathbf{\pm^{412}_{48}}$& \textbf{4.8} $\mathbf{\pm^{1.3}_{0.5}}$ & \textbf{71.0} & \textbf{71.5} & \textbf{93.6} & \textbf{10.40} & \textbf{3}\\
$\Delta t^{-10}$ & 124 $\pm^{165}_{155}$ & 259 $\pm^{441}_{89}$& 4.8 $\pm^{1.8}_{0.6}$ & 36.0 & 92.6 & 95.5 &  8.19 & 3\\
\hline

\sidehead{\textbf{2011 March 15: VLT/NaCo Ks}}
$\mathbf{\Delta t^{-1}}$ & \textbf{22} $\mathbf{\pm^{0.03}_{165}}$ & \textbf{358} $\mathbf{\pm^{342}_{8}}$& \textbf{4.2} $\mathbf{\pm^{0.6}_{0.4}}$ & \textbf{68.0} & \textbf{9.5} & \textbf{83.5} & \textbf{9.44} & \textbf{3}\\
$\Delta t^{-10}$ & 22 $\pm^{1}_{2}$ & 359 $\pm^{1}_{9}$& 4.1 $\pm^{0.6}_{0.3}$ & 31.7 & 5.7 & 82.9 & 9.31 & 3\\
\hline

\sidehead{\textbf{2013 March 27: VLT/NaCo H}}
$\mathbf{\Delta t^{-1}}$ & \textbf{-27} $\mathbf{\pm^{5}_{7}}$ & \textbf{31} $\mathbf{\pm^{9}_{11}}$& \textbf{3.8} $\mathbf{\pm^{0.8}_{0.8}}$ & \textbf{1} & $\mathbf{<0.1}$ & \textbf{84.1} & \textbf{32.01} & $\mathbf{>4}$\\
$\Delta t^{-10}$ & -28 $\pm^{5}_{7}$ & 29 $\pm^{18}_{9}$& 3.6 $\pm^{0.9}_{0.6}$ & 0.8 & $<0.1$ & 98.1 & 30.10 & $>4$
\enddata
\label{tbl-binfits_bad}
\tablenotetext{a}{Using distribution of noise simulation best fits}
\tablenotetext{b}{Using distribution of noise simulation F statistics}

\end{deluxetable*}
\renewcommand{\arraystretch}{1.0}

\subsection{2010 VLT/NaCo Ks Data - Published Non-detection}

Table \ref{tbl-binfits_bad} lists the results of binary fits to these data as well as the 2011 NaCo Ks and 2013 NaCo H band data sets. The $\Delta t^{-1}$ calibration strategy produced a best fit position angle of $122\pm^{66}_{137}$$^\circ$, separation of 268$\pm^{412}_{48}$ mas, and a contrast of $\Delta Ks = 4.8\pm^{1.3}_{0.5}$ mag. Appendix \ref{app-data_mod} shows these data alongside this work's best fit and the Huelamo et al. (2011) binary model.

\begin{figure*}[h!]
\epsscale{1.22}
\plotone{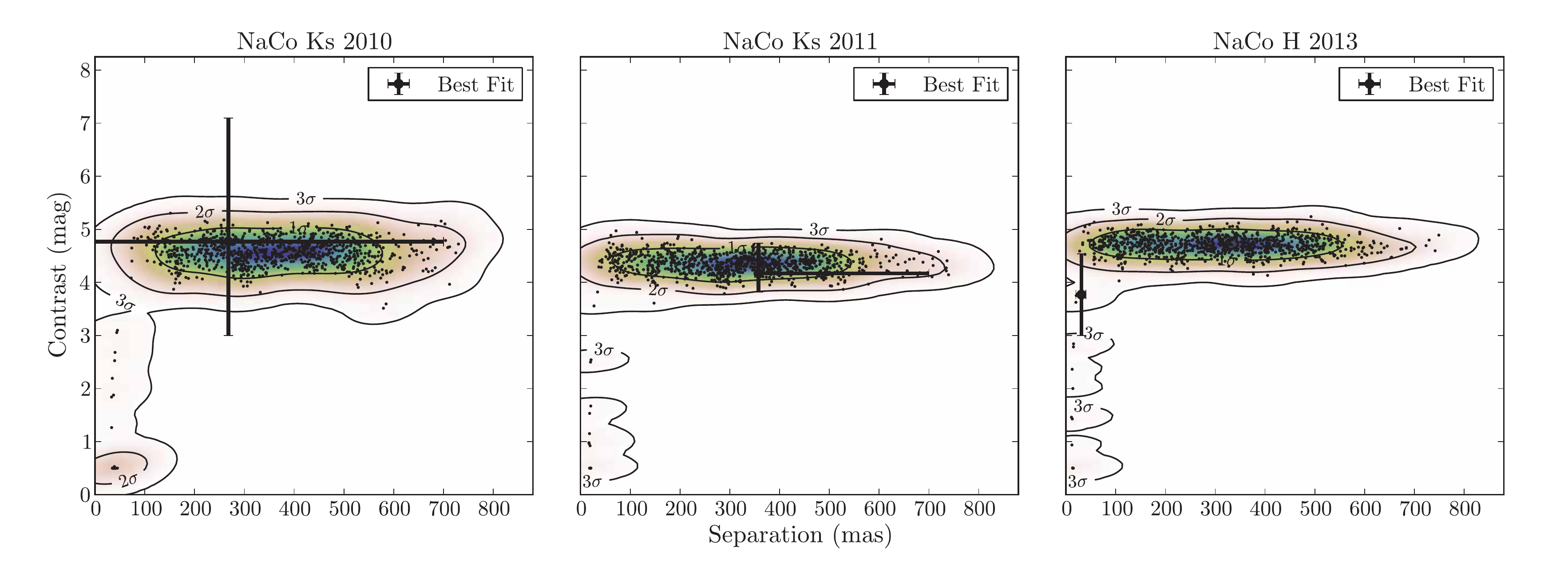}
\caption{Noise simulations for all H and Ks data sets. The scattered points show the best fits to 1000 noise realizations for each dataset, drawn from the Gaussian distributions shown in Figure \ref{fig-badcal}. The color shows the probability distribution interpolated from the best fits, while the contours indicate $1\sigma$, $2\sigma$, and $3\sigma$ confidence intervals. The bold point with error bars represents the best fit for each epoch.}\label{fig-bad_nsim}
\end{figure*}

\begin{figure*}[h!]
\epsscale{1.22}
\plotone{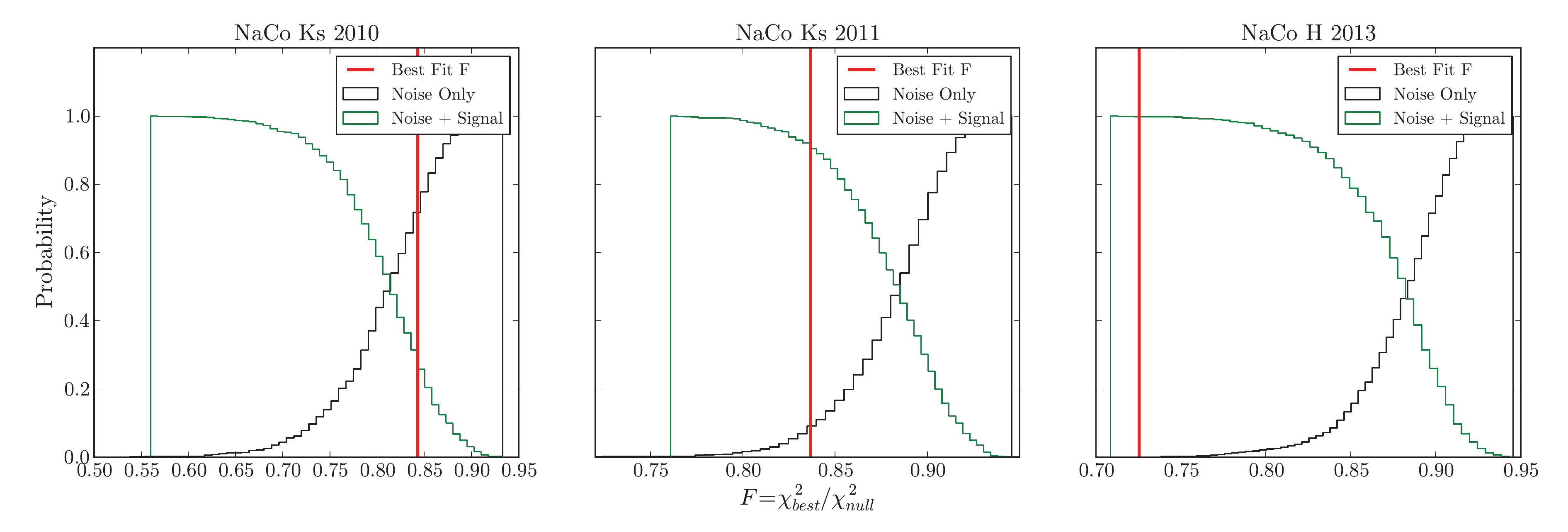}
\caption{False alarm testing for 2010 and 2011 Ks and 2013 H band observations. In each panel, the black, cumulative histogram shows the F statistics (best fit $\chi^2$ divided by null model $\chi^2$) for 1000 noise simulations. The intersection of the red, vertical line with the black histogram yields a false alarm probability. The green, cumulative histogram shows the F statistics from the 1000 noise + signal simulations carried out for each set of observations. The intersection of the red line with this histogram gives the fraction of noise + signal simulations that look less significant (higher F statistic) than the best fit.}\label{fig-bad_FAs}
\end{figure*}

For these data, there is a very high probability that the best fit resulted from noise, and that a companion signal would not have been recovered. The best fit to these data lies on a contour that encloses 29\% of the noise simulation results (see Figure \ref{fig-bad_nsim}). There is thus a 71\% chance that the best fit resulted from noise. The F statistics agree with this result; 71.5\% of the F statistics are lower than that measured for the best fit, $F = 0.843$ (see Figure \ref{fig-bad_FAs}). Both of these suggest that the fit to this dataset is consistent with noise.

For the 2010 and 2011 NaCo Ks and 2013 NaCo H band data sets, we estimate our type II errors following the procedure outlined in $\S$\ref{sec-sims}. As the input signal, we add companions with the same separation (62 mas) and contrast ($\Delta$ = 5.1) as the Huelamo et al. (2011) binary. To account for biases due to sky rotation coverage, we vary the input position angle of the companions. Of the 1000 noise + signal realizations, 936 resulted in erroneous best fits. This gives a 93.6\% chance that, with the noise present in the 2010 Ks data, we would not have recovered the companion signal had it been present. We thus cannot assign a non-detection to this dataset with a high degree of confidence.

\subsection{VLT/NaCo 2011 Ks Data}

The $\Delta t^{-1}$ calibration yielded a best fit position angle of $22\pm^{0.03}_{165}$$^\circ$, separation of $358\pm^{342}_{8}$ mas, and a contrast of $\Delta Ks = 4.2\pm^{0.6}_{0.45}$ mag. See Table \ref{tbl-binfits_bad} for $\Delta t^{-10}$ best fit parameters. Appendix \ref{app-data_mod} shows these data with this work's best fit and the Huelamo et al. (2011) binary model plotted.

Figure \ref{fig-bad_nsim} shows the results of noise simulations for the 2011 NaCo Ks data. The best fit falls on a contour that encloses 32\% of the noise simulations. This gives a 68\% chance that the best fit was caused by noise alone. The F statistics (see Figure \ref{fig-bad_FAs}), however, give a smaller false alarm probability; the best fit F statistic ($F = 0.837$) is higher than only 9.5\% of the simulated F statistics. This gives a 9.5\% chance that the best fit resulted from noise.

The type II error estimation gives an 83.5\% chance that the companion candidate signal would be lost under the noise. Of the 1000 realizations, 835 resulted in best fits that were inconsistent with the input signal. Thus, we cannot rule out the Huelamo et al. (2011) signal in these data.

\subsection{VLT/NaCo 2013 H Data}

The $\Delta t^{-1}$ best fit to these data has a position angle of -27$\pm^{5}_{7}$$^\circ$, separation of 29$\pm^{18}_{9}$, and contrast of 3.6$\pm^{0.9}_{0.6}$ mag. Table \ref{tbl-binfits_bad} lists the fit parameters for the $\Delta t^{-10}$ calibration. See Appendix \ref{app-data_mod} for a plot of these data, along with this work's best fit model and the Huelamo et al. (2011) companion model. 

Figure \ref{fig-bad_nsim} shows the results of noise and noise + signal simulations for the 2013 H band data. The best fit to these data, compared to the distribution of fits to noise, suggests that there is only a 1\% chance that the best fit is the result of noise. The F statistics also indicate that the false alarm probability is $<0.1\%$ (see Figure \ref{fig-bad_FAs}); the best fit F statistic ($F = 0.725$) is lower than all simulated F statistics. However, the kernel phase histogram (see Figure \ref{fig-badcal}) shows a significant number of outliers. The distribution is also not centered on zero. Both of these characteristics would cause the distribution of best fits and F statistics to underestimate the false alarm probability. 

Of the 1000 noise + signal simulations, we did not recover the input signal in 841 realizations. This gives an 84.1\% chance that we would have missed the companion signal, had it been present under the noise. Thus, we cannot rule out the Huelamo et al. (2011) signal with confidence using the H band data.

\section{Data and Model Plots}\label{app-data_mod}
Figures 14 - 22 show plots of kernel phase versus scan index, a proxy for sky rotation angle, for all observational epochs. In each plot, the grey points show the data and the red solid line marks this work's best fit. Figures 14 - 19 show data sets with low enough noise levels to include in our Discussion. Figure 19, which displays the MagAO/Clio2 data, shows the NaCo 2013 L$'$ best fit in green, since the two data sets were taken close in time to one another. All other figures in this section show the Huelamo et al. (2011) model in green. Figures 20 - 22 show data sets with high enough scatter to wash out signals of interest, and thus were not included in our Discussion.

\begin{figure*}[h!]
\epsscale{1.1}
\plotone{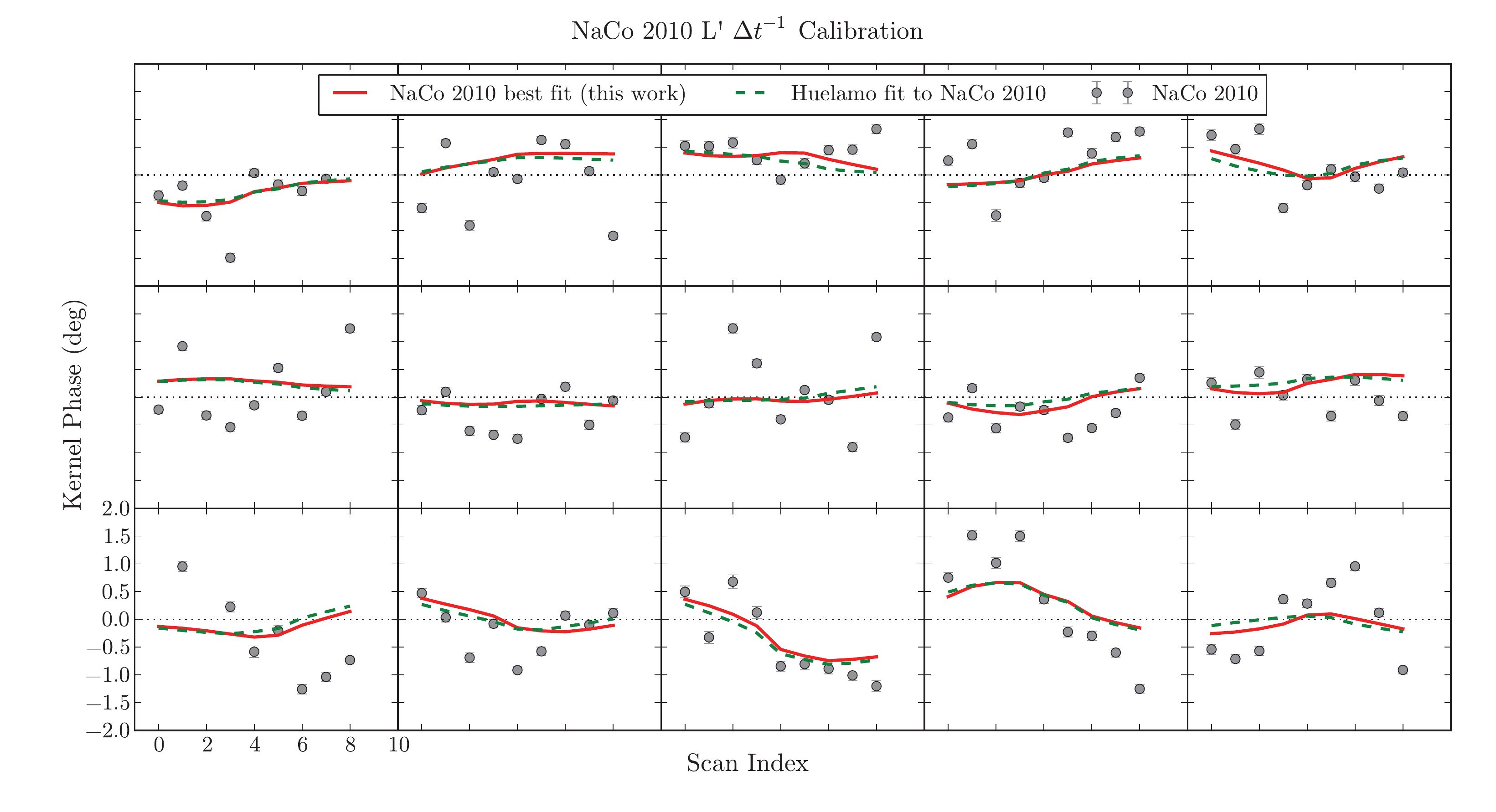}
\caption{Kernel phase data for 2010 NaCo L$'$ observations, shown with best fits from this work (red line) and Huelamo (dashed green line). Each subplot corresponds to a single linear combination of closure phases plotted against scan index (a proxy for sky rotation angle). The error bars plotted are unscaled, while our parameter constraints are derived using error bars scaled such that the reduced $\chi^2$ of the best fit model is equal to 1. The Huelamo et al. (2011) model is allowed within $1\sigma$ of this work's best fit.}
\label{fig-compfits_n10L}
\end{figure*}

\begin{figure*}[h!]
\epsscale{1.1}
\plotone{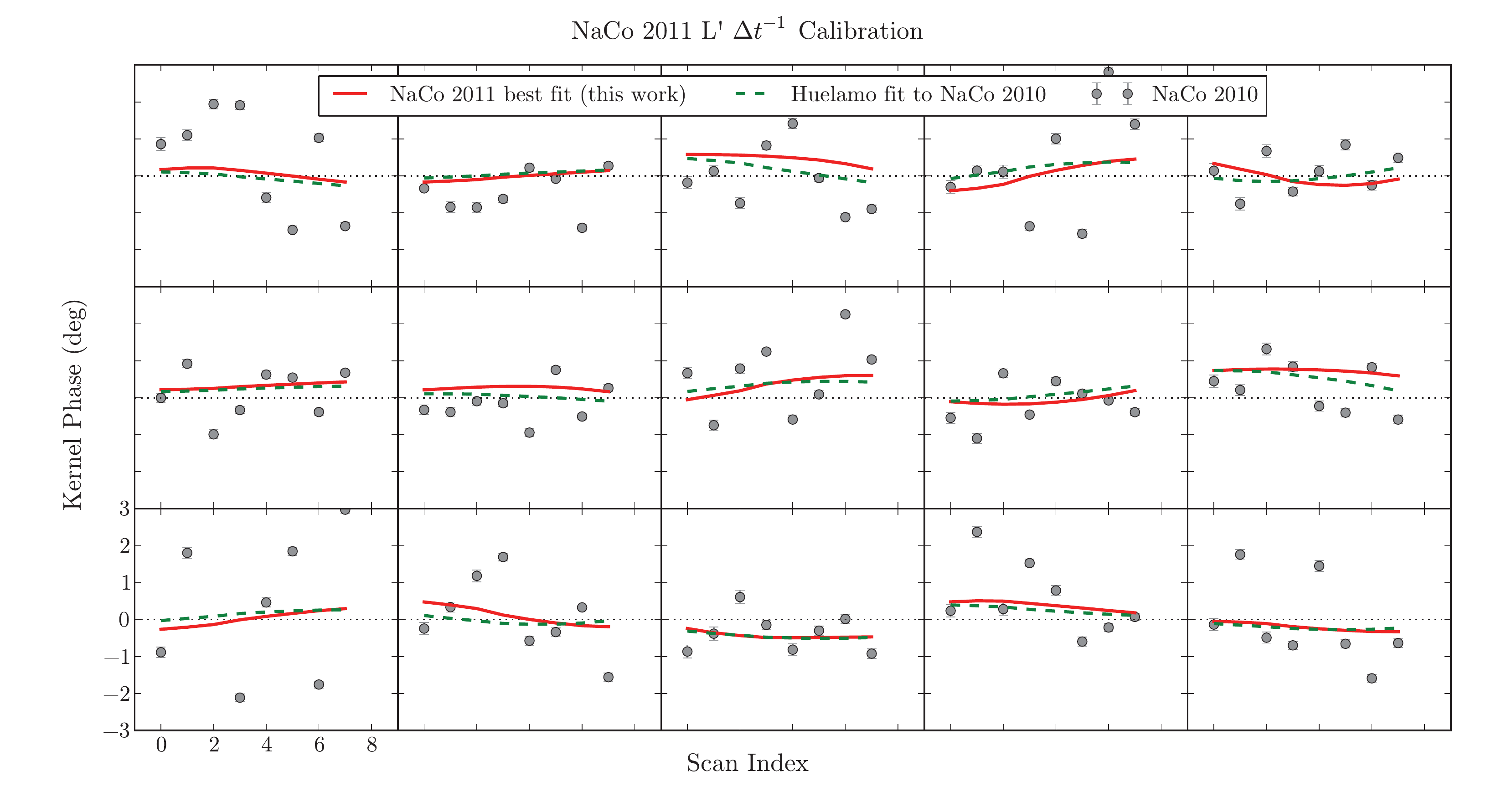}
\caption{Kernel phase data for 2011 NaCo L$'$ observations, shown with best fits from this work (red line) and Huelamo 2010 (dashed green line). Each subplot corresponds to a single linear combination of closure phases plotted against scan index (a proxy for sky rotation angle). The error bars plotted are unscaled, while our parameter constraints are derived using error bars scaled such that the reduced $\chi^2$ of the best fit model is equal to 1. The Huelamo et al. (2011) model is allowed within $1\sigma$ of this work's best fit.}
\label{fig-compfits_n11L}
\end{figure*}

\begin{figure*}[h!]
\epsscale{1.2}
\plotone{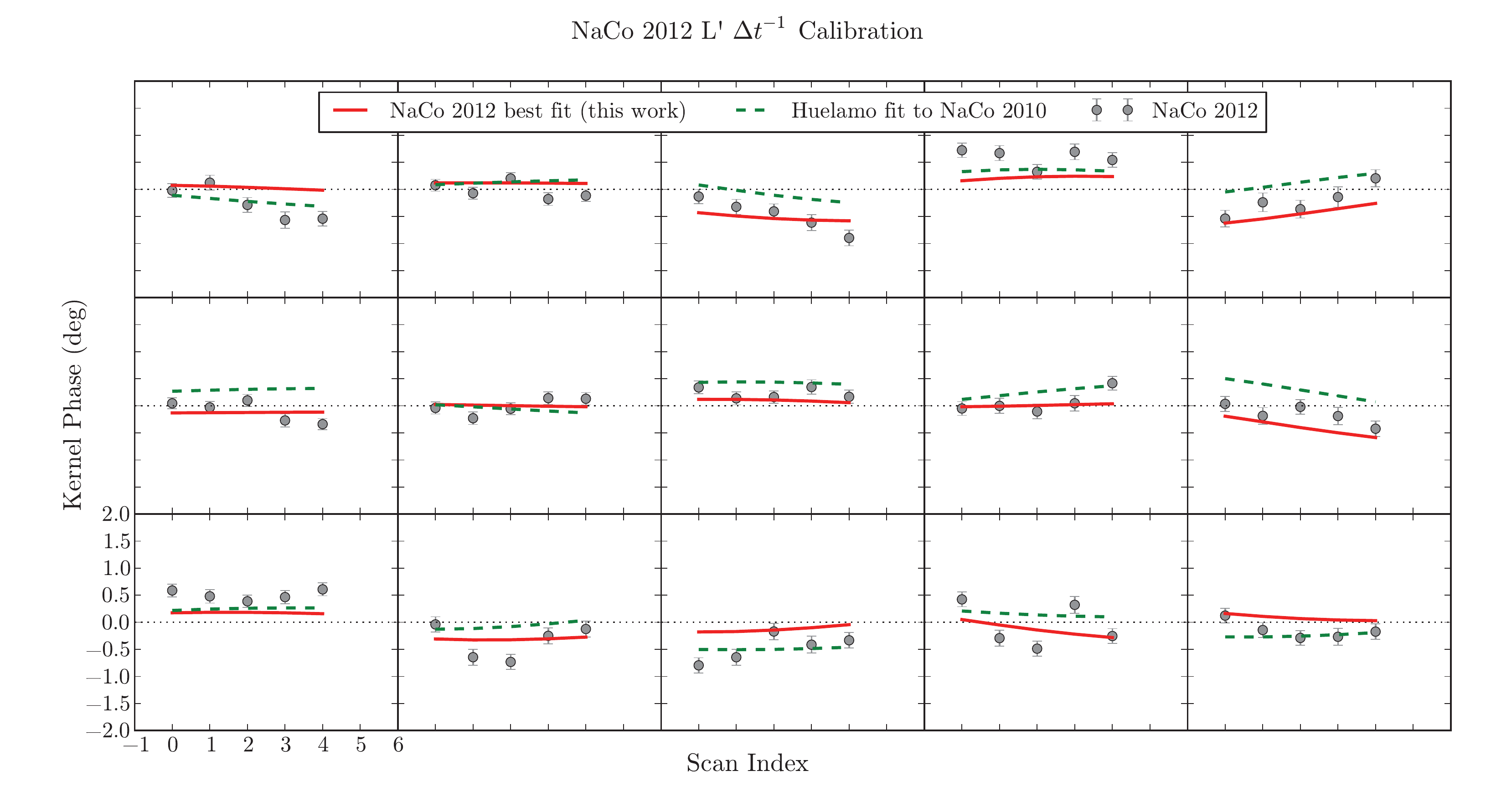}
\caption{Kernel phase data for 2012 NaCo L$'$ observations, shown with best fits from this work (red line) and Huelamo (dashed green line). Each subplot corresponds to a single linear combination of closure phases plotted against scan index (a proxy for sky rotation angle). The error bars plotted are unscaled, while our parameter constraints are derived using error bars scaled such that the reduced $\chi^2$ of the best fit model is equal to 1.}
\label{fig-compfits_n12L}
\end{figure*}

\begin{figure*}[h!]
\epsscale{1.1}
\plotone{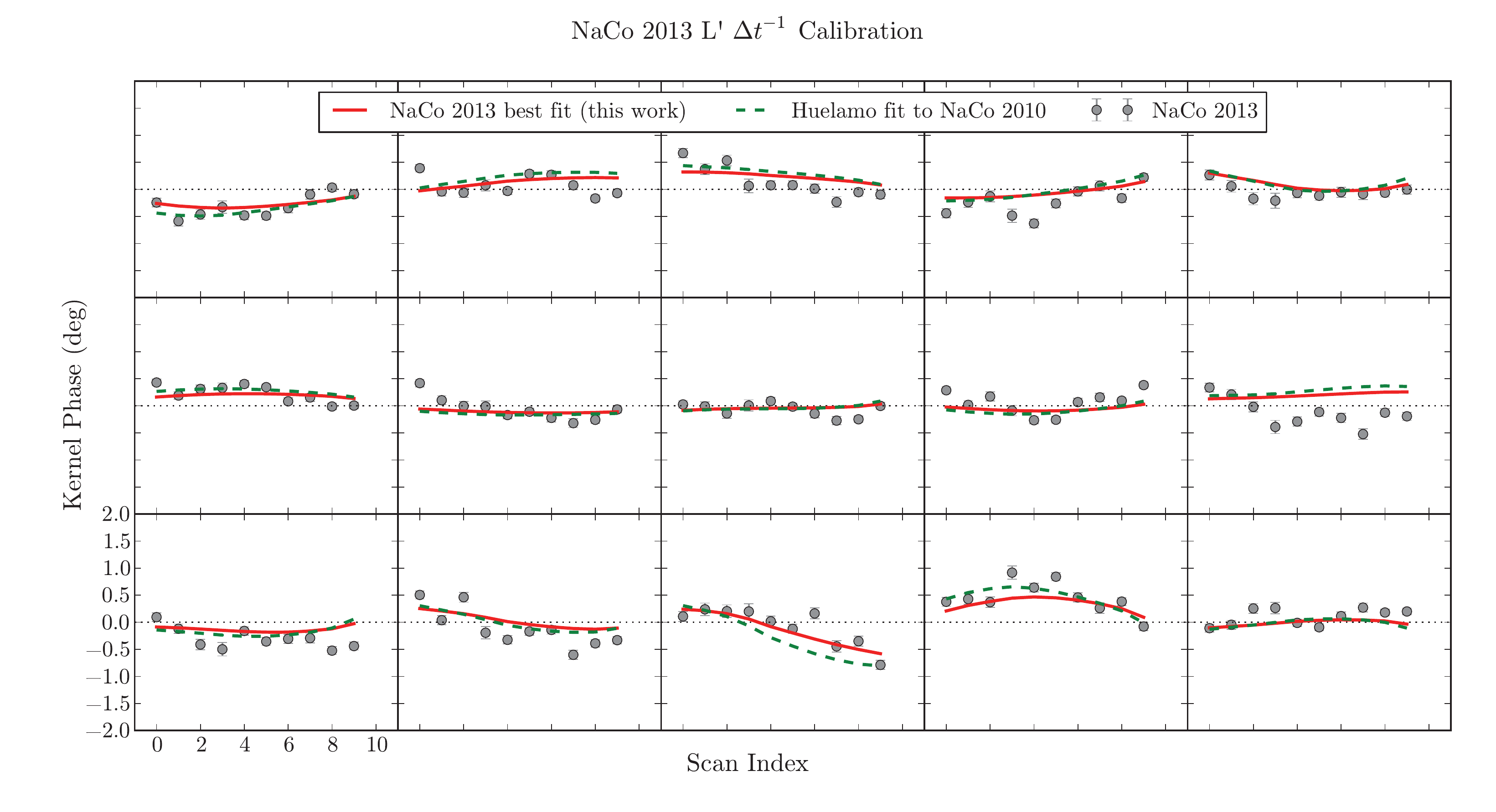}
\caption{Kernel phase data for 2013 NaCo L$'$ observations, shown with best fits from this work (red line) and Huelamo (dashed green line). Each subplot corresponds to a single linear combination of closure phases plotted against scan index (a proxy for sky rotation angle). The error bars plotted are unscaled, while our parameter constraints are derived using error bars scaled such that the reduced $\chi^2$ of the best fit model is equal to 1.}
\label{fig-compfits_n13L}
\end{figure*}

\begin{figure*}[h!]
\epsscale{1.1}
\plotone{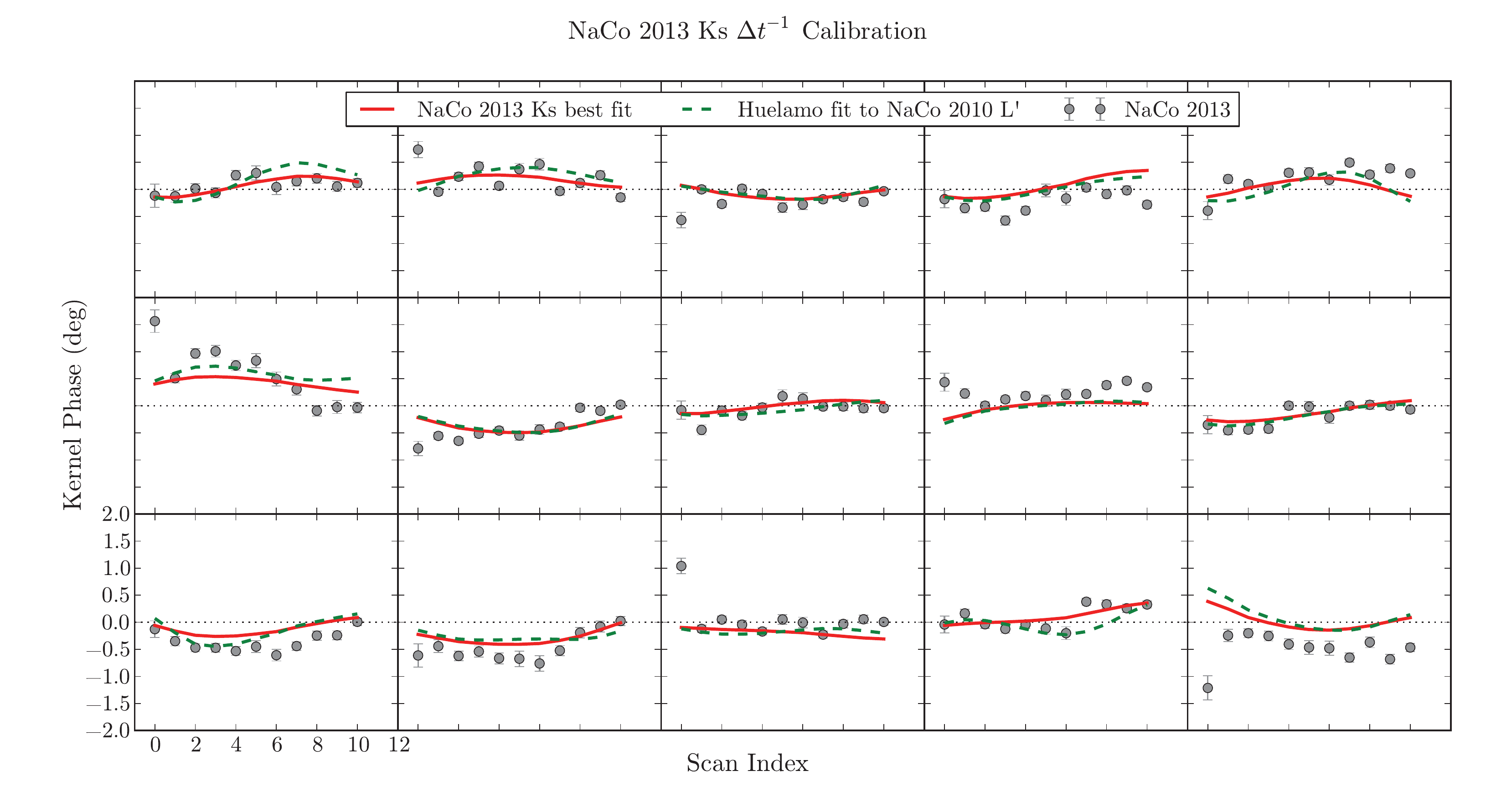}
\caption{Kernel phase data for 2013 NaCo Ks observations, shown with best fits from this work (red line) and Huelamo (dashed green line). Each subplot corresponds to a single linear combination of closure phases plotted against scan index (a proxy for sky rotation angle). The error bars plotted are unscaled, while our parameter constraints are derived using error bars scaled such that the reduced $\chi^2$ of the best fit model is equal to 1.}
\label{fig-compfits_n13K}
\end{figure*}

\begin{figure*}[h!]
\epsscale{1.1}
\plotone{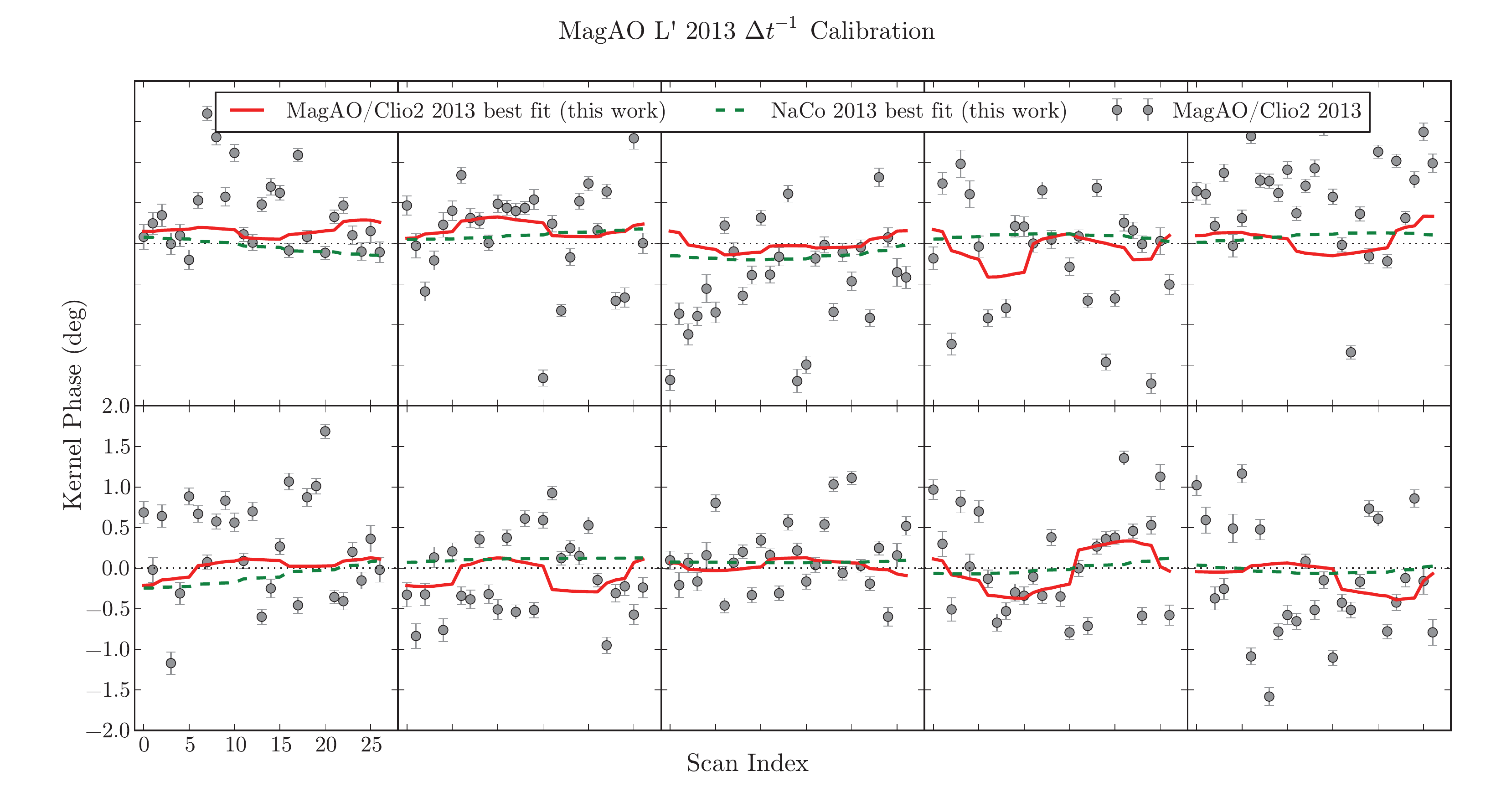}
\caption{Kernel phase data for 2013 MagAO L$'$ observations, shown with best fits from this work (red line) and the NaCo 2013 L$'$ best fit (dashed green line). Each subplot corresponds to a single linear combination of closure phases plotted against scan index (a proxy for sky rotation angle). The error bars plotted are unscaled, while our parameter constraints are derived using error bars scaled such that the reduced $\chi^2$ of the best fit model is equal to 1.}
\label{fig-compfits_m13L}
\end{figure*}

\begin{figure*}[h!]
\epsscale{1.1}
\plotone{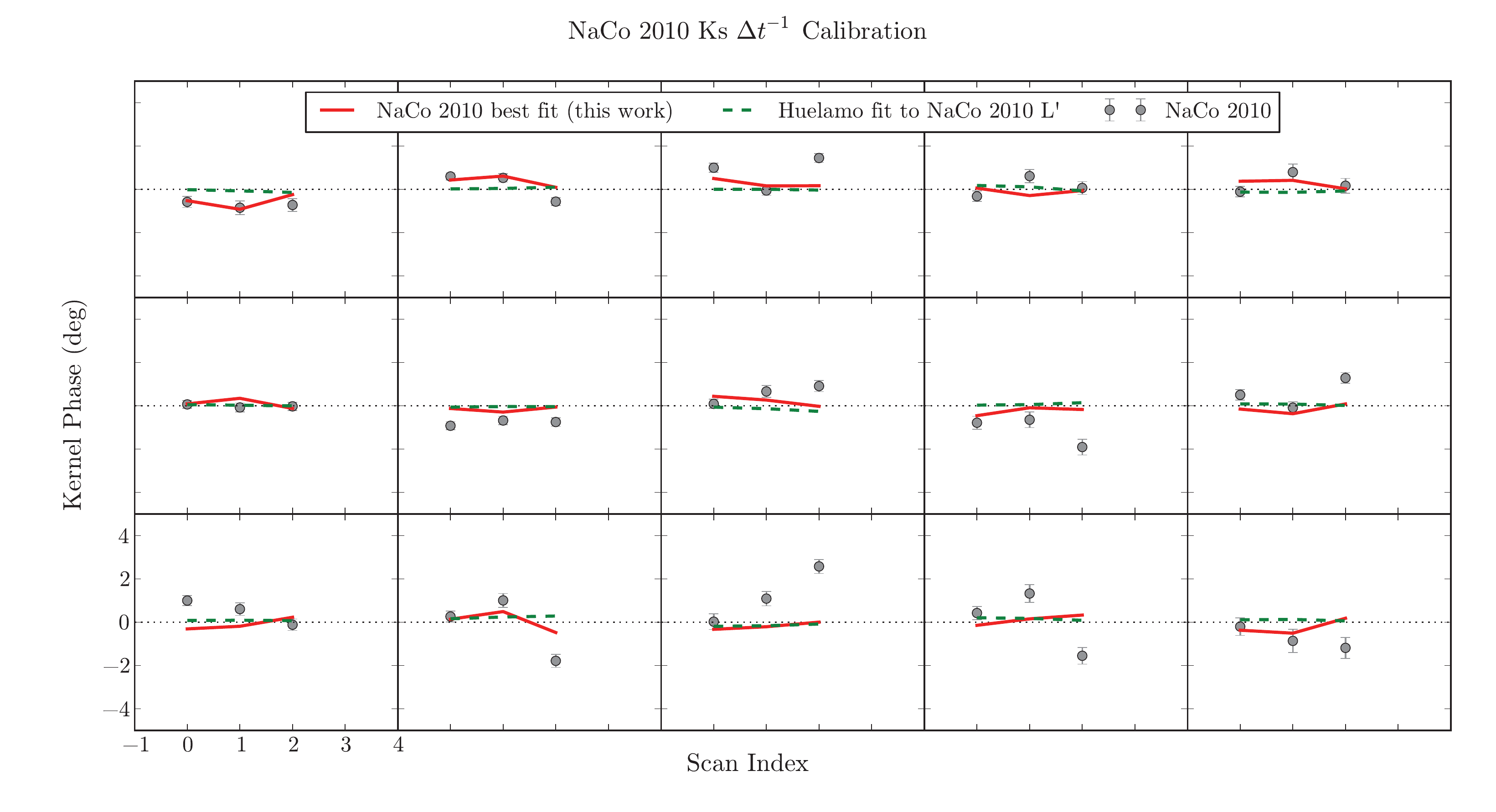}
\caption{Kernel phase data for 2010 NaCo Ks observations, shown with best fits from this work (red line) and Huelamo (dashed green line). Each subplot corresponds to a single linear combination of closure phases plotted against scan index (a proxy for sky rotation angle). The error bars plotted are unscaled, while our parameter constraints are derived using error bars scaled such that the reduced $\chi^2$ of the best fit model is equal to 1.}
\label{fig-compfits_n10K}
\end{figure*}

\begin{figure*}[h!]
\epsscale{1.1}
\plotone{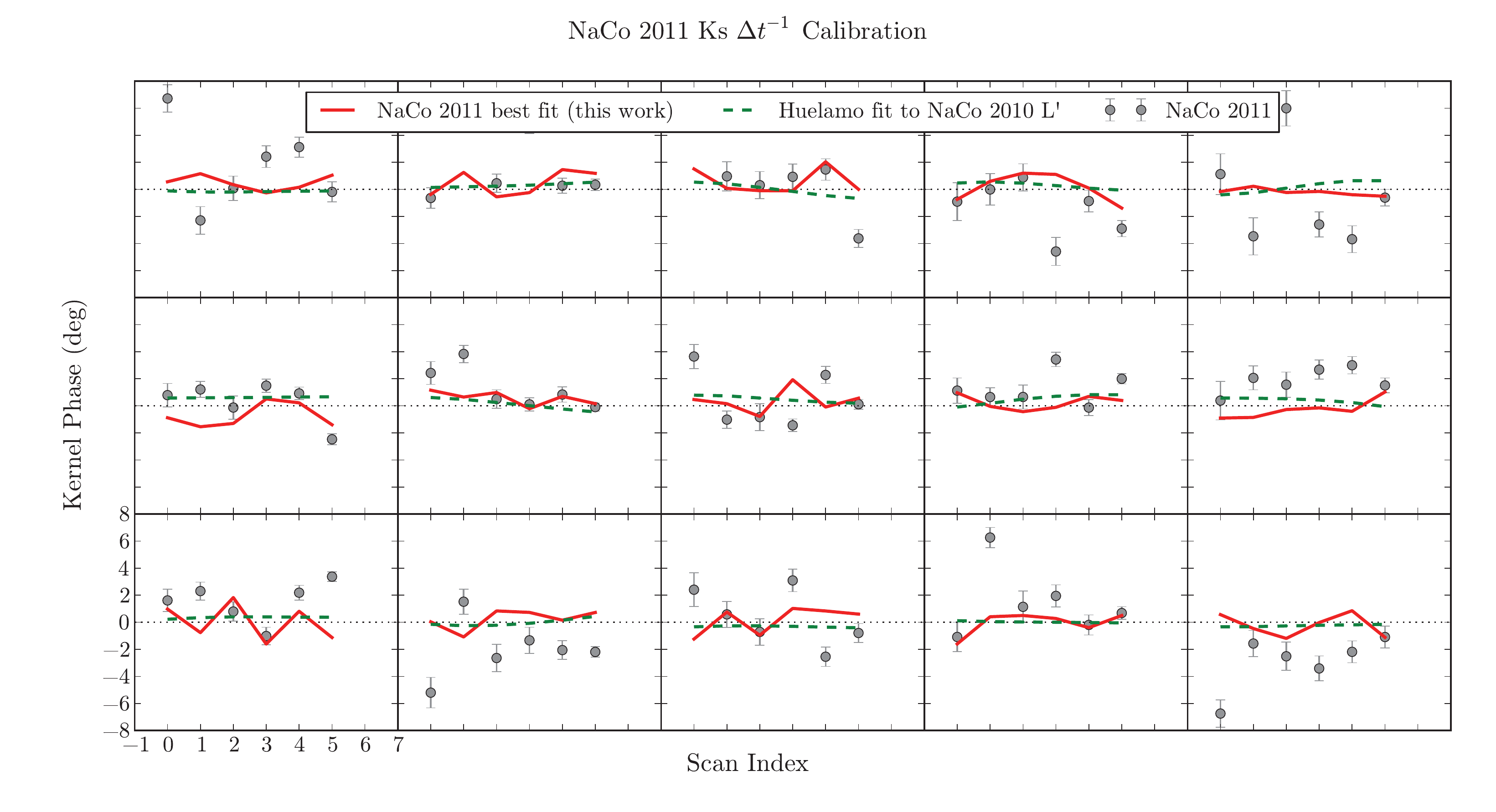}
\caption{Kernel phase data for 2011 NaCo Ks observations, shown with best fits from this work (red line) and Huelamo (dashed green line). Each subplot corresponds to a single linear combination of closure phases plotted against scan index (a proxy for sky rotation angle). The error bars plotted are unscaled, while our parameter constraints are derived using error bars scaled such that the reduced $\chi^2$ of the best fit model is equal to 1.}
\label{fig-compfits_n11K}
\end{figure*}

\begin{figure*}[h!]
\epsscale{1.1}
\plotone{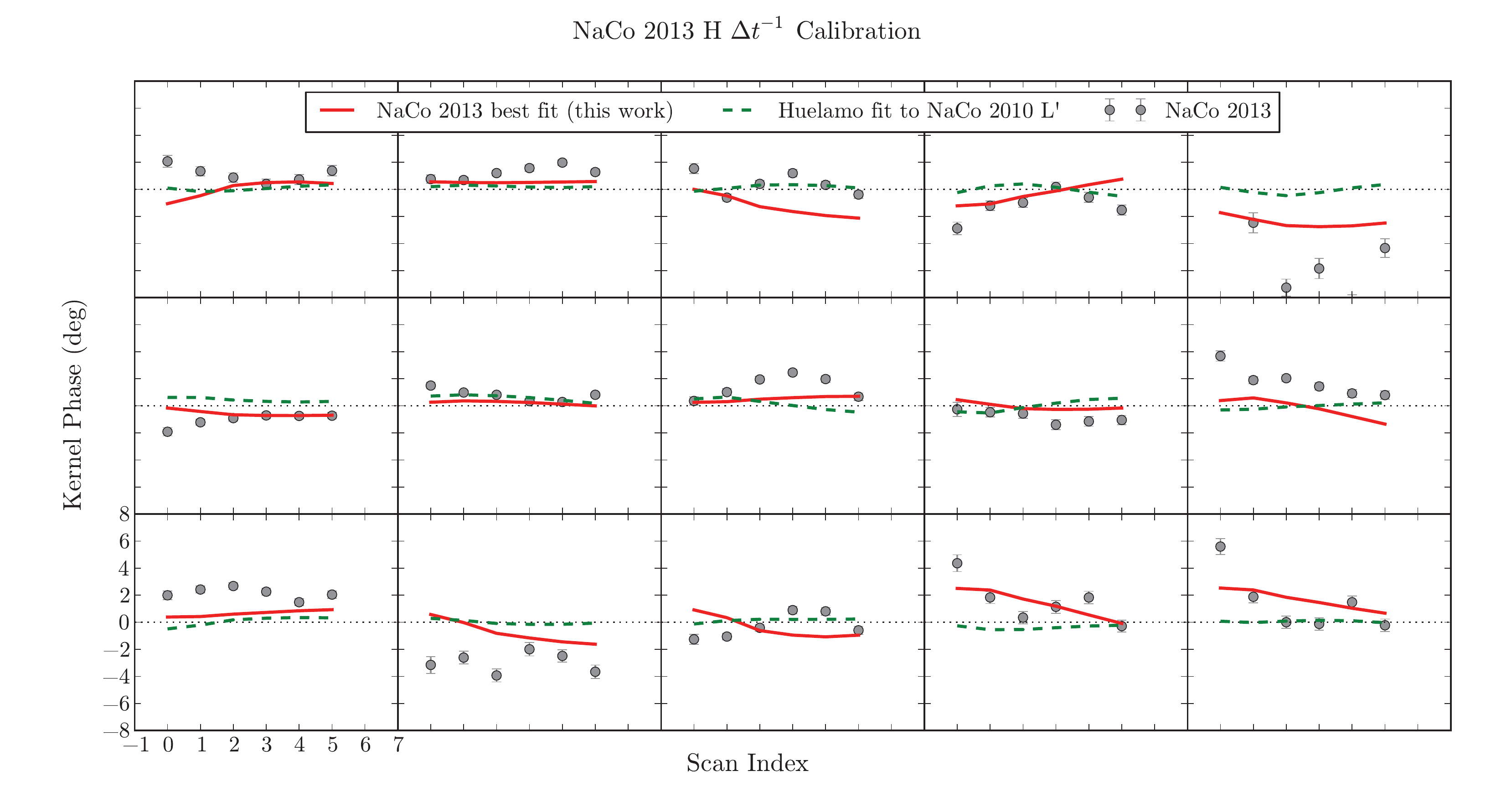}
\caption{Kernel phase data for 2013 NaCo H observations, shown with best fits from this work (red line) and Huelamo (dashed green line). Each subplot corresponds to a single linear combination of closure phases plotted against scan index (a proxy for sky rotation angle). The error bars plotted are unscaled, while our parameter constraints are derived using error bars scaled such that the reduced $\chi^2$ of the best fit model is equal to 1.}
\label{fig-compfits_n13H}
\end{figure*}

\section{Kernel Phase Projection}\label{app-kpproj}
Our kernel phase projection is similar to that in \citet{m10}. We begin with the matrix \textbf{A}, which describes the ways in which N$_{a}$ apertures ($\phi$) are combined to yield N$_{p}$ phases ($\Phi$):
\begin{equation}
\Phi = \textbf{A} \cdot \phi.
\end{equation}
This equation can be modified for observations of a source with intrinsic signal, assuming that the source phase simply adds to the instrumental phase:
\begin{equation}
\Phi = \textbf{A} \cdot \phi + \Phi_0.
\end{equation}
In order to eliminate the instrumental phase, we are searching for a matrix, \textbf{K}, such that:
\begin{equation}
\textbf{K} \cdot \textbf{A} = \textbf{0}.
\end{equation}
We can use singular value decomposition to find \textbf{K}.  We decompose \textbf{A$^T$} in the following way:
\begin{equation}
\textbf{A}^T = \textbf{U} \cdot \textbf{W} \cdot \textbf{V}^T
\end{equation}
where \textbf{U} is an N$_a$ $\times$ N$_p$ column-orthogonal matrix, \textbf{W} is an N$_p$ $\times$ N$_p$ diagonal matrix with either positive or zero elements, and \textbf{V} is an N$_p$ $\times$ N$_p$ orthogonal matrix.  The columns of \textbf{V} corresponding to zero \textbf{W}-values are filled into the rows of \textbf{K}. 
We then build \textbf{K} from a matrix, \textbf{T}, which describes how to combine phases into closure phases ($\Phi_{cp}$).
\begin{equation}
\Phi_{cp} = \textbf{T} \cdot \textbf{A} \cdot \phi + \textbf{T} \cdot \Phi_0 = \textbf{T} \cdot \Phi_0
\end{equation}
We find \textbf{B}, such that:
\begin{equation}
\textbf{B} \cdot \textbf{T} = \textbf{K}
\label{eq:BtoK}
\end{equation}
and
\begin{equation}
\Phi_{k} = \textbf{B} \cdot \textbf{T} \cdot \textbf{A} \cdot \phi + \textbf{B} \cdot \textbf{T} \cdot \Phi_0 = \textbf{K} \cdot \Phi_0.
\label{eq:BTAeq0}
\end{equation}

In the following equations, for any matrix \textbf{M}, \textbf{M}$_{right}^{-1}$ and \textbf{M}$_{left}^{-1}$ represent the right and left generalized inverses, respectively.  These are used to invert non-square matrices; a right inverse is required when a matrix has full row rank but does not have full column rank, and a left inverse when a matrix has full column rank but does not have full row rank.
\begin{equation}
 \textbf{M}_{right}^{-1} = \textbf{M}^T \cdot (\textbf{M} \cdot \textbf{M}^T)^{-1} 
\end{equation}
\begin{equation}
 \textbf{M}_{left}^{-1} = (\textbf{M}^T \cdot \textbf{M})^{-1} \cdot \textbf{M}^T 
\end{equation}
\textbf{K} has the dimensions N$_{k}$ $\times$ N$_{p}$ where N$_{k}$ is the number of (linearly independent) kernel phases and N$_{p}$ the number of Fourier phases (N$_k$ $<$ N$_{p}$). \textbf{K} has a right inverse.
\begin{equation}
\textbf{B} \cdot \textbf{T} \cdot \textbf{K}_{right}^{-1} = \textbf{K} \cdot \textbf{K}_{right}^{-1} = \textbf{I}.
\label{eq:binv}
\end{equation}

Since \textbf{B} has the dimensions N$_{k}$ $\times$ N$_{cp}$, (N$_k$ $<$ N$_{cp}$), \textbf{B} has a right inverse, which, according to \ref{eq:binv}, is the following:
\begin{equation}
\textbf{B}_{right}^{-1} = \textbf{T} \cdot \textbf{K}_{right}^{-1}
\end{equation}
and 
\begin{equation}
\textbf{B} \cdot \textbf{B}_{right}^{-1} = \textbf{I}
\label{eq:ref}
\end{equation}
Furthermore, the left inverse of \textbf{B}$_{right}^{-1}$ can be calculated to find the \textbf{B} in Equation (\ref{eq:ref}).

So:
\begin{equation}
\textbf{B} = (\textbf{B}_{right}^{-1})_{left}^{-1} =  (\textbf{T} \cdot \textbf{K}_{right}^{-1})_{left}^{-1}
\end{equation}
This satisfies both Equations (\ref{eq:BtoK}) and (\ref{eq:BTAeq0}).
We find the kernel phase covariance matrix, \textbf{C}$_k$, using the closure phase covariance matrix \textbf{C}$_{cp}$ in the following way:
\begin{equation}
\textbf{C}_k = \textbf{B} \cdot \textbf{C}_{cp} \cdot  \textbf{B}^T.
\label{eq:Ck}
\end{equation}
We can take our kernel phase variances to be the diagonal entries of \textbf{C}$_k$.

\clearpage


\begin{thebibliography}{}
\bibitem[Alcala et al. (1993)]{alc93} Alcala, J.M., Covino, E., Franchini, M. et al. 1993, A\&A, 272, 225
\bibitem[Alexander \& Armitage (2007)]{alexandarm07} Alexander, R. D., \& Armitage, P. J. 2007, MNRAS, 375, 500
\bibitem[Alexander et al. (2006)]{alexander06} Alexander, R. D., Clarke, C. J., \& Pringle, J. E. 2006, MNRAS, 369, 229
\bibitem[Andrews et al. (2011)]{andrews11} Andrews, S. M., Wilner, D. J., Espaillat, C., et al. 2011, ApJ, 732, 42
\bibitem[Artymowicz \& Lubow (1994)]{arty94} Artymowicz, P. \& Lubow, S. H. 1994, ApJ, 421, 651
\bibitem[Bans \& Kšnigl (2012)]{bans12} Bans, A. \& Kšnigl, A. 2012, ApJ, 758, 100
\bibitem[Bernat (2012)]{bernatphd} Bernat, D. 2012, PhD thesis, Cornell Univ. 
\bibitem[Birnstiel et al. (2012)]{birnstiel12} Birnstiel, T., Andrews, S. M., \& Ercolano, B. 2012 A\&A, 544, A79
\bibitem[Brown et al. (2009)]{brown09} Brown, J.M., Blake, G.A., Dullemond, C.P., Wilner, D.J., \& Williams, J.P. 2009, ApJ, 704, 496
\bibitem[Brown et al. (2007)]{brown07} Brown, J. M., Blake, G. A., Dullemond, C. P., et al. 2007, ApJ, 664, L107 
\bibitem[Bryden et al. (1999)]{bryden99} Bryden, G., Chen, X., Lin, D. N. C., Nelson, R. P., \& Papaloizou, J. 1999, ApJ, 514, 344
\bibitem[Calvet et al. (2002)]{calvet02} Calvet, N., D'Alessio, P., Hartmann, L., et al. 2002, ApJ, 568, 1008
\bibitem[Calvet et al. (2005)]{calvet05} Calvet, N., D'Alessio, P., Watson, D. M., et al. 2005, ApJL, 630, L185
\bibitem[Cieza et al. (2013)]{cieza13} Cieza, L. A., Lacour, S., Schreiber, M. R., et al. 2013, ApJL, 762, L12 
\bibitem[Clarke et al. (2001)]{clarke01} Clarke, C. J., Gendrin, A., \& Sotomayor, M. 2001, MNRAS, 328, 485
\bibitem[Close et al. (2013)]{close13} Close, L. M., Males, J. R., Morzinski, K., et al. 2013, ApJ, 774, 94
\bibitem[Crida et al. (2006)]{crida06} Crida, A., Morbidelli, A., \& Masset, F. 2006, Icar, 181, 587
\bibitem[D' Alessio et al. (2006)]{dalessio06} D' Alessio, P., Calvet, N., Hartmann, L., Franco-Hernandez, R., \& Servin, H. 2006, ApJ, 638, 314
\bibitem[D' Alessio et al. (1998)]{dalessio98} D' Alessio, P., Canto, J., Calvet, N., \& Lizano, S. 1998, ApJ, 500, 411
\bibitem[Dong et al. (2012)]{dong12} Dong, R., Rafikov, R., Zhu, Z., et al. 2012, ApJ, 750, 161
\bibitem[Draine et al. (2006)]{draine06} Draine, B. T. 2006, ApJ, 636, 1114
\bibitem[Drake et al. (2009)]{drake09} Drake, J. J., Ercolano, B., Flaccomio, E., \& Micela, G. 2009, ApJL, 699, L35
\bibitem[Dullemond \& Dominik (2004)]{dand04} Dullemond, C. P. \& Dominik, C. 2004, A\&A, 417, 159
\bibitem[Dullemond \& Dominik (2005)]{dandd05} Dullemond, C. P. \& Dominik, C. 2005, A\&A, 434, 971
\bibitem[Dullemond \& Natta (2003)]{dandn03} Dullemond, C. P. \& Natta, A. 2003, A\&A, 408,161
\bibitem[Ercolano et al. (2008)]{ercolano08} Ercolano, B., Drake, J. J., Raymond, J. C., \& Clarke, C. J. 2008, ApJ, 688, 398
\bibitem[Espaillat et al. (2007a)]{esp07a} Espaillat, C., Calvet, N., D'Alessio, P., et al. 2007a, ApJL, 670, L135
\bibitem[Espaillat et al. (2007b)]{esp07b} Espaillat, C., Calvet, N., D'Alessio, P., et al. 2007b, ApJL, 664, L111
\bibitem[Evans et al. (2012)]{evans12} Evans, T. M., Ireland, M. J., Kraus, A. L. et al. 2012, ApJ, 744, 120
\bibitem[Fouchet et al. (2010)]{fouchet10} Fouchet, L., Gonzalez, J. F., \& Maddison, S. T. 2010, A\&A, 518, A16
\bibitem[Freed et al. (2004)]{freed04} Freed, M. Hinz, P., Meyer, M. R., Milton, N. M., \& Lloyd-Hart, M. 2004, Proc. SPIE, 5492, 1561
\bibitem[Furlan et al. (2007)]{furlan07} Furlan, E., Sargent, B., Calvet, N., et al. 2007, ApJ, 664, 1176
\bibitem[Huelamo et al. (2011)]{hu11} Huelamo, N., Lacour, S., \& Tuthill, P. et al. 2011, A\&A, 528, L7
\bibitem[Hughes et al. (2009)]{hughes09} Hughes, A.M., Andrews, S. M., Espaillat C., et al. 2009, ApJ, 698, 131
\bibitem[Inoue et al. (2008)]{inoue08} Inoue, A. K., Honda, M., Nakamoto, T., \& Oka, A. 2008, PASJ, 60, 557
\bibitem[Ireland (2013)]{i13} Ireland, M. J. 2013, MNRAS, 433, 1718
\bibitem[Ireland \& Kraus (2014)]{ik14} Ireland, M. J. \& Kraus, A., 2014., in IAU Symp. 299, Exploring the Formation and Evolution of Planetary Systems (Cambirdge: Cambridge Univ. Press), 199
\bibitem[Ireland \& Kraus (2008)]{ik08} Ireland, M. J. \& Kraus, A. L. 2008, ApJ, 678, L59
\bibitem[Ireland et al. (2006)]{macim06} Ireland, M. J., Monnier, J. D., \& Thureau, N. 2006. Proc. SPIE, 6268, 58
\bibitem[Isella et al. (2013)]{isella13} Isella, A., Perez, L. M., Carpenter, J. M. et al. 2013, ApJ, 775, 30
\bibitem[Kraus \& Ireland (2012)]{ki12} Kraus, A. L. \& Ireland, M. J. 2012, ApJ, 745, 5
\bibitem[Kraus et al. (2011)]{k11} Kraus, A. L., Ireland, M. J., Martinache, F., \& Hillenbrand, L.A. 2011, ApJ, 731, 8
\bibitem[Kraus et al. (2013)]{kraus13} Kraus, S., Ireland, M. J., Sitko, M. L., et al. 2013, ApJ, 768, 80
\bibitem[Lacour et al. (2011)]{lacour11} Lacour, S., Tuthill, P., Amico, P. et al. 2011, A\&A, 532, A72
\bibitem[Lafreniere et al. (2007)]{loci} Lafreniere, D., Marois, C., Doyon, R., Nadeau, D., \& Artigau, E. 2007, ApJ, 660, 770 
\bibitem[Lin \& Papaloizou (1993)]{landp93} Lin, D. N. C. \& Papaloizou, J. 1993, in Protostars and Planets III, ed. E. H. Levy \& J. L. Lunine (Tucson, AZ: Univ. Arizona Press), 749
\bibitem[Lin \& Papaloizou (1986)]{landp86} Lin, D. N. C. \& Papaloizou, J. 1986, ApJ, 309, 846
\bibitem[Marois et al. (2000)]{marois00} Marois, C., Doyon, R., Racine, R., \& Nadeau, D. 2000, PASP, 112, 91
\bibitem[Martinache (2010)]{m10} Martinache, F. 2010, ApJ, 724, 464
\bibitem[Merin et al. (2010)]{merin10} Merin, B., Brown, J. M., Oliveira, I., et al. 2010, ApJ, 718, 1200
\bibitem[Morzinski et al. (2014)]{morzinski14} Morzinski, K., Close, L. M., Males, J. R., et al. 2014, Proc. SPIE, 9148, 04
\bibitem[Muto et al. (2012)]{muto12} Muto, T., Grady, C. A., Hashimoto, J., et al. 2012, ApJL, 748, L22
\bibitem[Najita et al. (2007)]{najita07} Najita, J. R., Strom S. E., \& Muzerolle, J. 2007, MNRAS, 378, 369
\bibitem[Natta et al. (2007)]{natta06} Natta, A., Testi, L., Calvet, N., et al. 2007 in Protostars \& Planets V, ed. B. Reipurth, D. Jewitt, \& K. Keil, (Tucson, AZ: Univ. Arizona Press), 767
\bibitem[Olofsson et al. (2011)]{ol11} Olofsson, J., Benisty, M., Augereau, J.-C., et al. 2011, A\&A, 528, L6
\bibitem[Olofsson et al. (2013)]{olofsson13} Olofsson, J., Benisty, M., \& Le Bouquin, J. -B. 2013, A\&A, 552, A4
\bibitem[Owen et al. (2011)]{owen11} Owen J. E., Ercolano, B., \& Clarke, C. J. 2011, MNRAS, 412, 13
\bibitem[Owen et al. (2010)]{owen10} Owen J. E., Ercolano, B., Clarke, C. J., \& Alexander, R. D. 2010, MNRAS, 401, 1415
\bibitem[Penzen (1993)]{penzen93} Penzen, R. 1993, Proc. SPIE, 1946, 635
\bibitem[Perez et al. (2014)]{perez14} Perez, L. M., Isella, A., Carpenter, J. M., \& Chandler, C. J. 2014, ApJL, 783, L13
\bibitem[Pichardo et al. (2008)]{pichardo08} Pichardo, B., Sparke, L. S., \& Aguilar, L. A. 2008, MNRAS, 391, 815
\bibitem[Pollack et al. (1985)]{pollack85} Pollack, J. B,. McKay, C. P., \& Christofferson, B. M. 1985, Icar, 64, 471
\bibitem[Press et al. (1992)]{numrecipes} Press, W. H., Teukolsky, S. A., Vetterling, W. T., \& Flannery, B. P. 1992, Numerical Recipes in C (New York: Cambridge Univ. Press)
\bibitem[Protassov \& van Dyk (2002)]{protassov02} Protassov, R. \& van Dyk, D. A. 2002, ApJ, 571, 545
\bibitem[Robitaille (2011)]{robitaille11} Robitaille, T. P. 2011, A\&A, 536, A79
\bibitem[Rousset et al. (2003)]{rousset03} Rousset, G., Lacombe, F., Puget, P. et al. 2003, Proc. SPIE, 4839, 140
\bibitem[Schisano et al. (2009)]{schisano09} Schisano, E., Covino, E. Alcala, J. M. et al. 2009, A\&A, 501, 1013
\bibitem[Schlosman \& Begelman (1987)]{sandb87}
\bibitem[Shakura \& Sunyaev (1973)]{sands73} Shakura, N. I., \& Sunyaev, R. A. 1973, A\&A, 24, 337
\bibitem[Sitko et al. (2008)]{sitko08} Sitko, M. L., Carpenter, W. J., Kimes, R. L., et al. 2008, ApJ, 678, 1010
\bibitem[Sivanandam et al. (2006)]{sivana06} Sivanadam, S., Hinz, P., Heinze, A. N., et al. 2006, Proc. SPIE, 6269
\bibitem[Sivia \& Skilling (2006)]{sands06} Sivia, D., \& Skilling, J. 2006, Data Analysis: A Bayesian Tutorial (Oxford: Oxford Univ. Press)
\bibitem[Strom et al. (1989)]{strom89} Strom, K.M., Strom, S. E., Edwards, S., Cabrit, S., \& Skrutskie, M. F. 1989, AJ, 97, 1451
\bibitem[Tanaka et al. (2005)]{tanaka05}Tanaka, H., Himeno, Y., \& Shigru, I. 2005 ApJ, 625, 414
\bibitem[Torres et al. (2008)]{torres08} Torres, C. A. O., Quast, G. R., Melo, C. H. F., Sterzik, M. F. 2008, Handbook of Star Forming Regions, Volume II: The Southern Sky ASP Monograph Publications, Vol. ed. Bo Reipurth, 757
\bibitem[Uchida et al. (2004)]{uchida04} Uchida, K. I., Calvet, N., Hartmann, L., et al. 2004 ApJS, 154, 439
\bibitem[Windmark et al. (2012)]{windmark12} Windmark, F., Birnstiel, T., GŸttler, C. et al. 2012, A\&A, 540, A73
\bibitem[Wisniewski et al. (2008)]{wisnie08} Wisniewski, J. P., Clampin, M., Grady, C. A., et al. 2008, ApJ, 682, 548
\end{thebibliography}
\end{document}